\documentclass[longauth]{aa} 

\usepackage{amssymb}
\usepackage{graphicx}
\usepackage{txfonts}
\usepackage{xcolor} 
\usepackage[flushleft]{threeparttable} 
\usepackage{subcaption} 
\usepackage{hyperref}
\usepackage{placeins}
\defcitealias{Samland2017}{SAM17}

\begin{document} 

\newcommand{\SubItem}[1]{
    {\setlength\itemindent{15pt} \item[-] #1}
}

   \title{Revisiting the atmosphere of the exoplanet 51 Eridani b with VLT/SPHERE}

   \author{S.~B.~Brown-Sevilla
          \inst{1}\fnmsep\thanks{Member of the International Max-Planck Research School for Astronomy and Cosmic Physics at the University of Heidelberg (IMPRS-HD), Germany}
          \and A.-L. Maire \inst{2,3}
          \and P. Molli\`ere\inst{1}
          \and M. Samland \inst{1}
          \and M. Feldt \inst{1}
          \and W. Brandner \inst{1}
          \and Th. Henning \inst{1}
          \and R. Gratton \inst{4} 
          \and M. Janson \inst{5} 
          \and T. Stolker \inst{6} 
          \and J. Hagelberg \inst{7} 
          \and A. Zurlo \inst{8,9,10} 
          \and F. Cantalloube \inst{8} 
          \and A. Boccaletti \inst{11} 
          \and M. Bonnefoy \inst{2} 
          \and G. Chauvin \inst{2,12} 
          \and S. Desidera \inst{4} 
          \and V. D'Orazi \inst{4}
          \and A.-M. Lagrange \inst{2,11} 
          \and M. Langlois \inst{13} 
          \and F. Menard \inst{2} 
          \and D. Mesa \inst{4} 
          \and M. Meyer \inst{14} 
          \and A. Pavlov \inst{1} 
          \and C. Petit \inst{15} 
          \and S. Rochat \inst{2} 
          \and D. Rouan \inst{9} 
          \and T. Schmidt \inst{9} 
          \and A. Vigan \inst{8} 
          \and L. Weber \inst{7}. 
          }

   \institute{Max Planck Institute for Astronomy, K\"onigstuhl 17, 69117, Heidelberg, Germany\\
              \email{brown@mpia.de}
         \and
             Universit\'e Grenoble Alpes, CNRS, IPAG, 38000 Grenoble, France 
        \and
            STAR Institute, Universit\'e de Li\`ege, All\'ee du Six Ao\^ut 19c, B-4000 Li\`ege, Belgium 
        \and
            INAF -- Osservatorio Astronomico di Padova, Vicolo dell'Osservatorio 5, 35122 Padova, Italy
        \and
            Department of Astronomy, Stockholm University, 10691, Stockholm, Sweden
        \and
            Leiden Observatory, Leiden University, P.O. Box 9513, 2300 RALeiden, The Netherlands
        \and
            Geneva Observatory, University of Geneva, Chemin des Mailettes 51, 1290 Versoix, Switzerland
        \and
            Aix Marseille Univ, CNRS, CNES, LAM, Marseille, France
        \and
            N\'ucleo de Astronom\'ia, Facultad de Ingenier\'ia y Ciencias, Universidad Diego Portales, Av. Ej\'ercito 441, Santiago, Chile
        \and
            Escuela de Ingenier\'ia Industrial, Facultad de Ingenier\'ia y Ciencias, Universidad Diego Portales, Av. Ejercito 441, Santiago, Chile
        \and
            LESIA, Observatoire de Paris, Université PSL, CNRS, Sorbonne Universit\'e, Universit\'e de Paris, 5 place Jules Janssen, 92195 Meudon, France
        \and    
            Unidad Mixta Internacional Franco-Chilena de Astronomía, CNRS/INSU UMI 3386 and Departamento de Astronomía, Universidad de Chile, Casilla 36-D, Santiago, Chile
        \and
            CRAL, UMR 5574, CNRS, Universit\'e de Lyon, École Normale Sup\'erieure de Lyon, 46 Allée d’Italie, 69364 Lyon Cedex 07, France
        \and
            European Southern Observatory (ESO), Karl-Schwarzschild-Str. 2, 85748 Garching, Germany
        \and
            DOTA, ONERA, Universit\'e Paris Saclay, F-91123, Palaiseau France\\
             }

   \date{Received August 30, 2022; accepted November 21, 2022}

 
  \abstract
   {}
   {We aim to better constrain the atmospheric properties of the directly imaged exoplanet 51~Eri~b by using a retrieval approach on higher signal-to-noise data than previously reported. In this context, we also compare the results of using an atmospheric retrieval vs a self-consistent model to fit atmospheric parameters.}
   {We apply the radiative transfer code \texttt{petitRADTRANS} to our near-infrared SPHERE observations of 51~Eri~b in order to retrieve its atmospheric parameters. Additionally, we attempt to reproduce previous results with the retrieval approach and compare the results to self-consistent models using as priors the best-fit parameters from the retrieval.}
   {We present a higher signal-to-noise $YH$ spectrum of the planet and revised $K1K2$ photometry (M$_{K1} = 15.11 \pm 0.04$ mag, M$_{K2} = 17.11 \pm 0.38$ mag). The best-fit parameters obtained using an atmospheric retrieval differ from previous results using self-consistent models. In general, we find that our solutions tend towards cloud-free atmospheres (e.g. log $\tau_{\rm clouds} = -5.20 \pm 1.44$). For our ``nominal'' model with new data, we find a lower metallicity ([Fe/H] $= 0.26\pm$0.30 dex) and C/O ratio ($0.38\pm0.09$), and a slightly higher effective temperature (T$_{\rm{eff}} = 807\pm$45 K) than previous studies. The surface gravity (log $g = 4.05\pm0.37$) is in agreement with the reported values in the literature within uncertainties. We estimate the mass of the planet to be between 2 and 4 M$_{\rm{Jup}}$. When comparing with self-consistent models, we encounter a known correlation between the presence of clouds and the shape of the \textit{P-T} profiles.}
   {Our findings support the idea that results from atmospheric retrievals should not be discussed in isolation, but rather along with self-consistent temperature structures obtained using the retrieval's best-fit parameters. This, along with observations at longer wavelengths, might help to better characterise the atmospheres and determine their degree of cloudiness.}


   \maketitle
%

\section{Introduction}

The development of adaptive optics (AO) in recent years has allowed ground-based instruments such as the Gemini Planet Imager \citep[GPI;][]{Macintosh2014} and the Spectro-Polarimetric High-contrast Exoplanet REsearch instrument \citep[SPHERE;][]{Beuzit2019} to detect tens of substellar companions \citep[e.g.,][]{Bowler2017, Janson2019, Bohn2020}. Direct imaging allows to perform spectroscopic observations of the companions to probe the properties and composition of their atmospheres. High-contrast imaging is sensitive to the thermal near infrared (NIR) emission of recently formed giant planets and brown dwarfs. The young age of these objects makes them excellent targets for testing planet formation theories \citep[e.g.,][]{Spiegel2012, Mordasini2009, Mordasini2009b}, by comparing their luminosity with evolutionary track predictions for models of differing initial entropy, such as hot- or cold-start models \citep[e.g.,][]{Allard2012}. In addition, advancements in the treatment of clouds in atmospheric models, allow to better characterise the detected companions \citep[e.g.,][]{Baudino2015, Molliere2020, Carrion-Gonzalez2020}.

\object{51 Eridani b} is the first discovered planet by the GPI exoplanet survey \citep{Macintosh2015}. It was first characterised using both $J-$ and $H-$band spectra from GPI, and $L_P-$band photometry from Keck/NIRC2. This young giant planet shows strong methane spectral signatures, an unusual feature in most directly imaged exoplanets. The planet orbits \object{51 Eridani A}, a young F0IV star member of the $\beta$ Pictoris moving group \citep{Zuckerman2001, Bell2015}. The latest estimate for the isochronal age of the system from \textit{Gaia} EDR3 is $\sim$10~Myr \citep{Lee2022}, a much younger age than the commonly adopted $\sim$20~Myr. Using photometry from the Transiting Exoplanet Survey Satellite (TESS), \cite{Sepulveda2022} recently determined that 51~Eri is a $\gamma$ Doradus pulsator with a core rotation period of 0.9$_{-0.1}^{+0.3}$~days. The star is part of a hierarchical triple system, along with the M-dwarf binary GJ 3305AB, separated by $\sim$2000 au \citep{Feigelson2006, Kasper2007}. 51 Eri is located at 29.90$\pm$0.06 pc, as derived from the precise parallax measurement by the \textit{Gaia} mission \citep{Gaia2021}. From 24 $\mu$m \textit{Spitzer} observations, 51 Eri is known to have an infrared (IR) excess \citep{Rebull2008}, and a debris disk was detected using \textit{Herschel} observations at 70 and 100 $\mu$m with a very low IR fractional luminosity of $L_{\rm{IR}}/L_* = 2.3 \times 10^{-6}$ and a lower limit on the inner radius of 82 au \citep{Riviere-Marichalar2014}. From WISE observations, \cite{Patel2014} report a warm disk (T$\sim$180 K) at a radius of 5.5 au assuming blackbody radiation. Therefore, it is likely that the architecture of 51 Eri could resemble that of our Solar System with a two-belt debris disk.

The planet 51~Eri~b was confirmed to be bound to the system in a follow-up paper by \cite{DeRosa2015}. In \cite{DeRosa2020}, the authors presented a revised version of the orbital parameters of the planet (inclination $i = 136\substack{+10 \\ -11}$ deg, semi-major axis $a = 11.1\substack{+4.2 \\ -1.3}$ au, and orbital period $P = 28.1\substack{+17.2 \\ -4.9}$ yr assuming a mass of 1.75 M$_\odot$ for the host star) from GPI observations, which they found to be consistent with the parameters derived by \cite{Maire2019} from SPHERE observations. The inclination found in these studies suggests that the planet is not coplanar with the binary \object{GJ 3305AB} \citep[$i = 92.1 \pm 0.2$,][]{Montet2015}. The discovery paper used the luminosity and age of the system to derive a mass estimate for 51~Eri~b of $2-12\,\mathrm{M_{Jup}}$ for a cold-start formation scenario \citep{Macintosh2015}. Later estimates reported by \cite{Samland2017} (hereafter SAM17) ranged from 2.4 to 5 $\mathrm{M_{Jup}}$ for hot-start, and up to 12~$\mathrm{M_{Jup}}$ for warm-start models. \cite{DeRosa2020} showed that a dynamical mass measurement of the planet with \textit{Gaia} may be possible if the mass of the planet is $M\,\gtrsim 4\,\mathrm{M_{Jup}}$. Recently, \cite{Muller2021} reported mass and metallicity estimates for 51~Eri~b derived from synthetic cooling tracks and the planet's luminosity. Assuming an age range of 17–23 Myr and a hot-start formation scenario, they obtain a mass of $M = 2.3\,\mathrm{M_{Jup}}$ and a metallicity of [Fe/H] $=$ 0.11. Another study by \cite{Dupuy2022} presented an upper limit for the mass of the planet of $M < 11\,\mathrm{M_{Jup}}$ at 2$\sigma$ using cross-calibration of \textit{Hipparcos} and \textit{Gaia} EDR3 astrometry. They also revise the luminosity of 51~Eri~b using a photometric approach and find log$(L_{\rm{bol}}/L_\odot) = -5.5 \pm 0.2$~dex. Additionally, they derived a lower limit on the initial specific entropy of the planet which rules out cold-start formation scenarios.
 
In addition to the discovery paper, there have been two atmospheric analyses to characterise 51~Eri~b. By combining SPHERE/IFS $YJ$ and $YH$, and GPI $H$ spectra, along with photometry from SPHERE (broad-band $H$, $H23$ and $K12$) and Keck/NIRC2 ($L_P$), \citetalias{Samland2017} found the atmosphere to be cloudy. They report $T_{\rm{eff}} = 760 \pm 20$ K, $R = 1.11^{+0.16}_{-0.14}\,\mathrm{R_{Jup}}$, log~$g = 4.26 \pm 0.25$ dex, [Fe/H] $= 1.0 \pm 0.1$ dex, and $f_{\rm{sed}} = 1.26^{+0.36}_{-0.29}$ for their best-fit model. On the other hand, \cite{Rajan2017} used updated GPI $J$ and $H$ spectra and updated Keck/NIRC2 $L_P$ photometry from the discovery paper, along with new Keck/NIRC2 $M_S$ photometry and determined the atmosphere to be partially cloudy. Their best-fit model yielded $T_{\rm{eff}} =$ 605-737 K, [Fe/H] = 1.0 and log~$g =$ 3.5-4.0 dex. The three studies made use of self-consistent models and differ mainly in the degree of cloudiness of the atmosphere.
   
In this work, we present new NIR spectro-photometric observations of 51~Eri~b obtained using VLT/SPHERE as part of the SHINE survey \citep{Desidera2021, Langlois2021, Vigan2021}. These observations were carried out as a follow-up to the ones presented in \cite{Samland2017} and have the highest signal-to-noise (S/N) achieved so far (S/N $\sim$ 23 for $K1$). We use the radiative transfer code \texttt{petitRADTRANS} to model the atmospheric spectrum of the planet. Additionally, we attempt to reproduce the results in \citetalias{Samland2017} using a retrieval approach.
   
The paper is structured as follows: in Sect.~\ref{sec:obs} we describe our spectro-photometric observations. In Sect.~\ref{sec:red} we detail our data reduction and spectrum extraction procedures, as well as the derived detection limits in planet contrast and mass. The description of the atmospheric retrieval runs is detailed in Sect.~\ref{sec:retr}. We present a detailed analysis of certain parameters of the planet and a discussion of the results in Sect.~\ref{sec:discussion}, and finally, we summarise our results in Sect.~\ref{sec:summary}.

\begin{table}
\caption[51 Eri parameters]{51 Eri physical parameters assumed in this study.} 
	\centering
	\begin{tabular}{ccc}
	\hline
    \hline
    	51 Eri & Value & Ref. \\
        \hline
        Spectral type  & F0IV & a \\
		Age  & 10 - 20 Myr & b, c \\ 
		Distance \textit{d} & 29.900 $\pm$ 0.067 pc & d \\ 
		Mass & 1.75 $\pm$ 0.05 M$_\odot$ & e \\ 
        Visual magnitude ($V$ band) & 5.200 $\pm$ 0.009 mag & f \\
        Fe/H & 0.07 - 0.11 & g \\
        \hline
        \hline
	\end{tabular}
	\begin{tablenotes}
	    \item {\footnotesize \textbf{Notes.} \hspace{1em} a) \cite{Abt1995}, b) \cite{Lee2022}, c) \cite{Macintosh2015}, d) \cite{Gaia2021}, e) \cite{Simon2011}, f) \cite{Hog2000}, g) \cite{Arentsen2019}}
	\end{tablenotes} 
\label{tab:pitch}
\end{table}

\section{Observations}
\label{sec:obs}

   \begin{figure}
   \centering
   \includegraphics[width=0.38\textwidth]{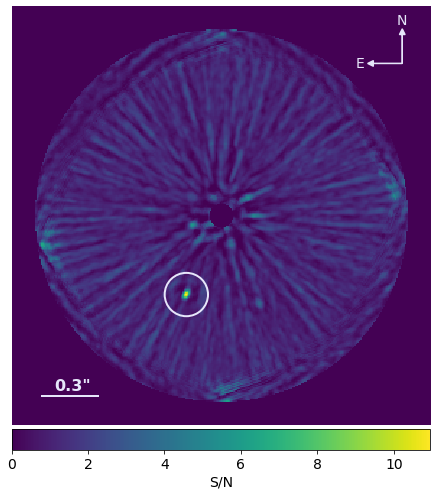}
   \caption{Median combined S/N detection map from ANDROMEDA from the SPHERE/IFS data. The circle indicates the position of 51~Eri~b. The azimuthal wings around the planet signal is the characteristic planet signature that ANDROMEDA is fitting for in ADI data.}
              \label{fig:snr_map}
    \end{figure}

   New data of 51 Eri were obtained with the VLT/SPHERE high-contrast instrument \citep{Beuzit2019} within the SHINE \citep[SpHere INfrared survey for Exoplanets,][]{Chauvin2017} Guaranteed Time Observations (GTO) program on the night of September 28, 2017.
   These observations were carried out in the IRDIS\_EXT mode, which combines IRDIS in dual-band imaging mode \citep[DBI;][]{Vigan2010} in the $K1K2$ ($K1=2.110\ \mu$m and $K2=2.251\ \mu$m) filters with IFS \citep{Claudi2008} in the $YH$ spectral bands (between 0.95 and 1.65 $\mu$m, with spectral resolution $R \sim$ 33). An apodized pupil Lyot coronagraph with a focal mask diameter of 185 milli-arcsec, was used for the observations \citep{Carbillet2011}. In order to reduce residual speckle noise, the observations were carried out close to meridian passage using the pupil stabilized mode, which allows the use of ADI post-processing \citep{Marois2006}.
   
   To calibrate the flux and center of the images, unsaturated non-coronagraphic images (hereafter referred to as the point spread function or PSF), as well as coronagraphic images with the deformable mirror (DM) waffle mode \citep[see][for more details on this mode]{Langlois2013} are acquired at the beginning and at the end of the observing sequence. The waffle mode generates four artificial replicas of the star in a ``cross'' pattern, commonly known as satellite spots. These spots are used to measure the star's position at the center of the pattern. In order to minimize the uncertainties in the frame centering and the astrometric error, and monitor the photometric stability throughout the sequence, the science frames were also obtained using this mode. 
   Finally, night-time sky images were acquired to estimate the background level in the science frames. The pixel scale and the True North (or north angle offset) were obtained using astrometric calibrators, as described in \cite{Maire2016}. The astrometric calibration of the data set was done using an observation of an astrometric field performed during the observing run. The usual calibration images (i.e. flat-field, bias, and spectral calibration) were obtained by the internal calibration hardware of the instrument during daytime after the observations. The observations were conducted in overall favorable conditions (see Table \ref{tab:obslog}), except for the presence of clouds near the end of the observing sequence.
   
   \begin{table}
   \centering
   \caption[Observing log]{Log of observations.} 
   \begin{threeparttable}
	\begin{tabular}{lc}
	\hline
    \hline
        UT date  & 28-09-2017 \\
        Observing mode & IRDIS\_EXT \\
        IRDIS filter & K1K2 \\
        IFS band & YH \\
        IRDIS DIT & 24 s \\
        IRDIS NDIT & 32 \\
        IFS DIT & 32 s \\
        IFS NDIT & $7\times 22$ \\
	    Field rotation & 44.1 deg \\
	    Strehl ratio & $0.85-0.91$ \\
	    Airmass (start/end) & $1.10-1.09$ \\
	    Seeing & $0.4-0.7$" \\
	    Coherence time ($\tau_0$) & $5-12$ ms \\
        \hline
 	\end{tabular}  
 	\begin{tablenotes}
	    \item {\footnotesize \textbf{Notes.} DIT stands for detector integration time. The Strehl ratio is measured at 1.6\,$\mu$m. The seeing and coherence time are measured at 0.5\,$\mu$m.} 
	\end{tablenotes}
   \end{threeparttable}
\label{tab:obslog}
\end{table}


\section{Data reduction and spectrum extraction}
\label{sec:red}

\subsection{Data reduction}

The data were reduced with the SPHERE Data Center pipeline \citep{Delorme2017}, using the SPHERE Data Reduction Handling (DRH) software \citep[version 15.0;][]{Pavlov2008}. This basic reduction consists in performing the usual sky background subtraction, flat fielding, bad-pixel identification and interpolation, star centering corrections and, for IFS, the calibration of the wavelengths and of the cross-talk between spectral channels. We then removed poor-quality frames where a significant drop in stellar flux is detected from the satellite spots photometry, because of interfering cirrus clouds near the end of the observing sequence, only the first 140 of the total of 154 IFS frames were used for the post-processing analysis. We also discarded the second PSF for the flux normalization and only used the first PSF frames. The conditions were very stable from the start of observations for the first PSF frames until the onset of cirrus clouds. Additionally, we tested different SPHERE data reduction recipes and pipelines, see Appendix~\ref{sec:data_reduction_sphere} for a detailed discussion on how they compare.

For the post-processing data analysis, we used the ANgular Differential OptiMal Exoplanet Detection Algorithm \citep[ANDROMEDA;][]{Cantalloube2015}, which utilises angular differential imaging (ADI) and an inverse problem approach based on a maximum likelihood estimator \citep{Mugnier2009} to search for companion candidates. It carries out a pair-wise subtraction of frames at different rotation angles and performs a cross-correlation of the signature that a point source would leave in the residual image \citep[see][for more details]{Cantalloube2015}. The main outputs from ANDROMEDA are 2D maps of: the estimated contrast of detected point sources at every location in the image; the corresponding standard deviation on this contrast; and the S/N ratio obtained dividing the contrast by the standard deviation. 
Figure~\ref{fig:snr_map} shows the resulting S/N detection map from ANDROMEDA for our IFS data. Other than 51~Eri~b, we detect no additional point sources.

Additionally, we performed the reduction with TRAP, a temporal, non-local systematics modelling algorithm to look for point sources at small separations \citep[see][]{Samland2021}. Regarding the S/N ratio, with ANDROMEDA we get S/N = 23.31 for $K1$, and S/N = 2.82 for $K2$, while with TRAP we get S/N = 18.03 for $K1$, and S/N = 3.52 for $K2$. In both cases, there is an improvement in the S/N ratio from \citetalias{Samland2017} (7.46 and 1.26 for $K1$ and $K2$, respectively using ANDROMEDA). On the other hand, the contrast limits are improved with the TRAP reduction as discussed below in Sect.~\ref{sec:det_lim}. Due to the higher S/N ratio achieved in $K1$, and to be consistent throughout the paper, we decided to use the results of ANDROMEDA for the following steps.

\subsection{Spectrum calibration}

To construct the spectrum of 51~Eri~b, we multiplied the planet contrast at each wavelength from ANDROMEDA by a template spectrum of the host star. This template spectrum was obtained as follows: we used a model stellar spectrum from the BT-NextGen library \citep{Allard2012} with $T_\mathrm{eff} = 7200$ K, log $g = 4.0$ dex, [Fe/H] $= 0.0$ dex, and no alpha enhancement (overabundance of He with respect to metallicity, [$\alpha$/Fe]), since these parameters are the closest to the ones determined from high-resolution spectra for 51~Eri ($T_\mathrm{eff} = 7256$ K, log $g = 4.13$ dex, and [Fe/H] $= 0.0$ dex; \citealt{Prugniel2007}). Then we fit this model spectrum to the spectral energy distribution (SED) of the star using the $\chi ^2$ minimization in the Virtual Observatory SED Analyzer \citep{Bayo2008} to obtain the flux scaling factor to account for the distance of 51~Eri. The SED was built with photometry from Tycho $B_{\rm{T}}$, $V_{\rm{T}}$ \citep{Hoeg1997}, WISE W3 \citep{Cutri2013}, Johnson $U$, $V$, $B$ \citep{Mermilliod2006}, and IRAS 12 $\mu$m \citep{Helou1988}. Finally, we scaled the model spectrum to the resolution of our IFS data using the python function \texttt{SpectRes}\footnote{\url{https://spectres.readthedocs.io/en/latest/}}.

A similar procedure along with the respective transmission curves for the filters was used to obtain updated IRDIS photometry for $K1$ and $K2$. Since the planet is not detected significantly in the $K2$ filter, we applied forced photometry with ANDROMEDA. This consists in performing a photometric measurement in the $K2$ images at the position of the planet in the $K1$ frames. To estimate the uncertainties in the flux of both the spectrum and the photometry, we propagated the standard deviation obtained with ANDROMEDA for the IFS and IRDIS data, respectively. Our results are shown in Table~\ref{tab:phot}.

\begin{figure}
 \centering
 \includegraphics[width=0.48\textwidth]{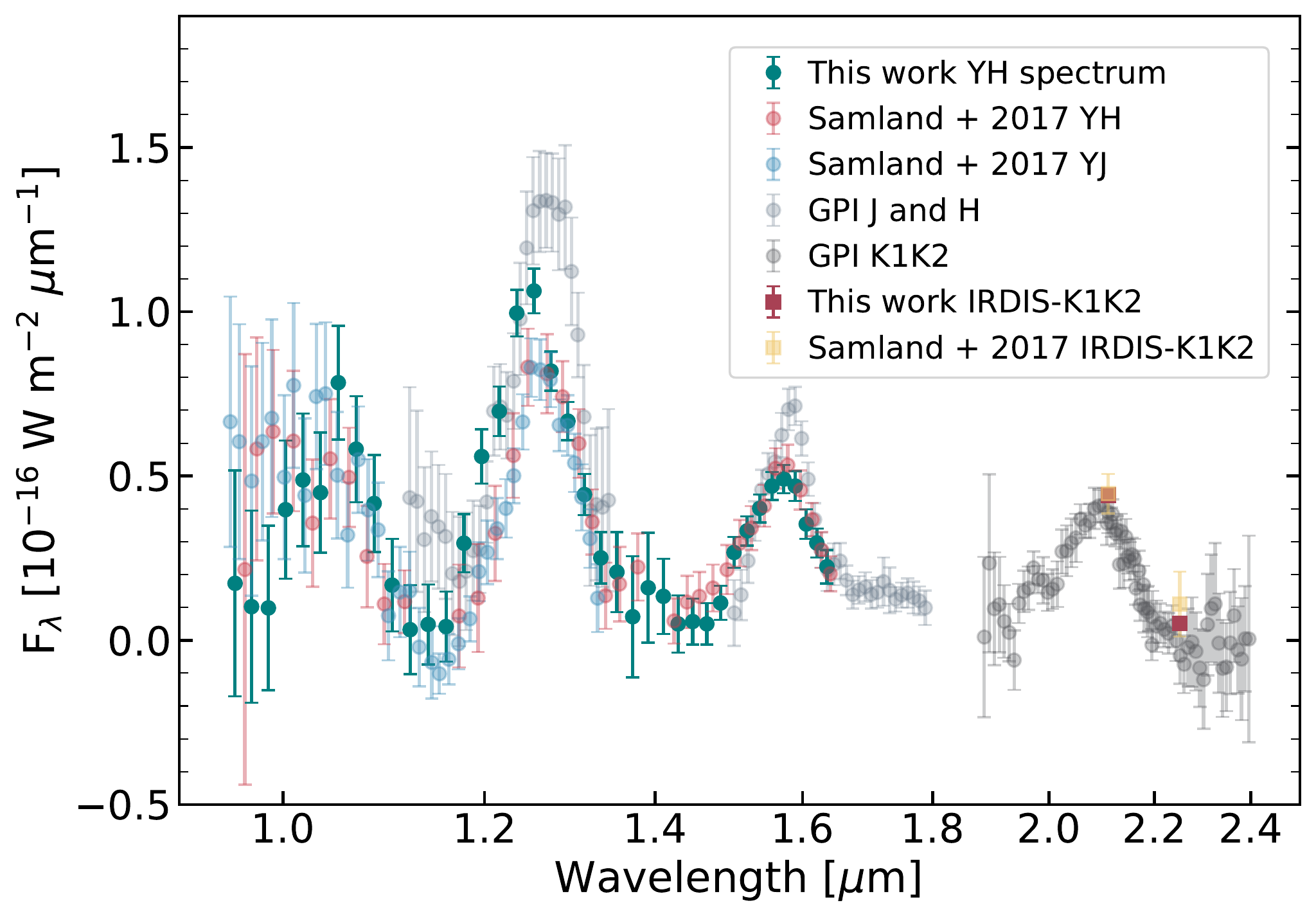}
 \caption{The newly obtained 51~Eri~b $YH$ spectrum and $K1K2$ photometry. Overplotted for comparison the $YH$ spectrum and $K1K2$ photometry from \cite{Samland2017}, as well as the updated GPI $J$ and $H$ spectra and the $K1$ and $K2$ spectra from \cite{Rajan2017}.}
    \label{fig:spec}
\end{figure}

\begin{figure}
 \centering
 \includegraphics[width=0.49\textwidth]{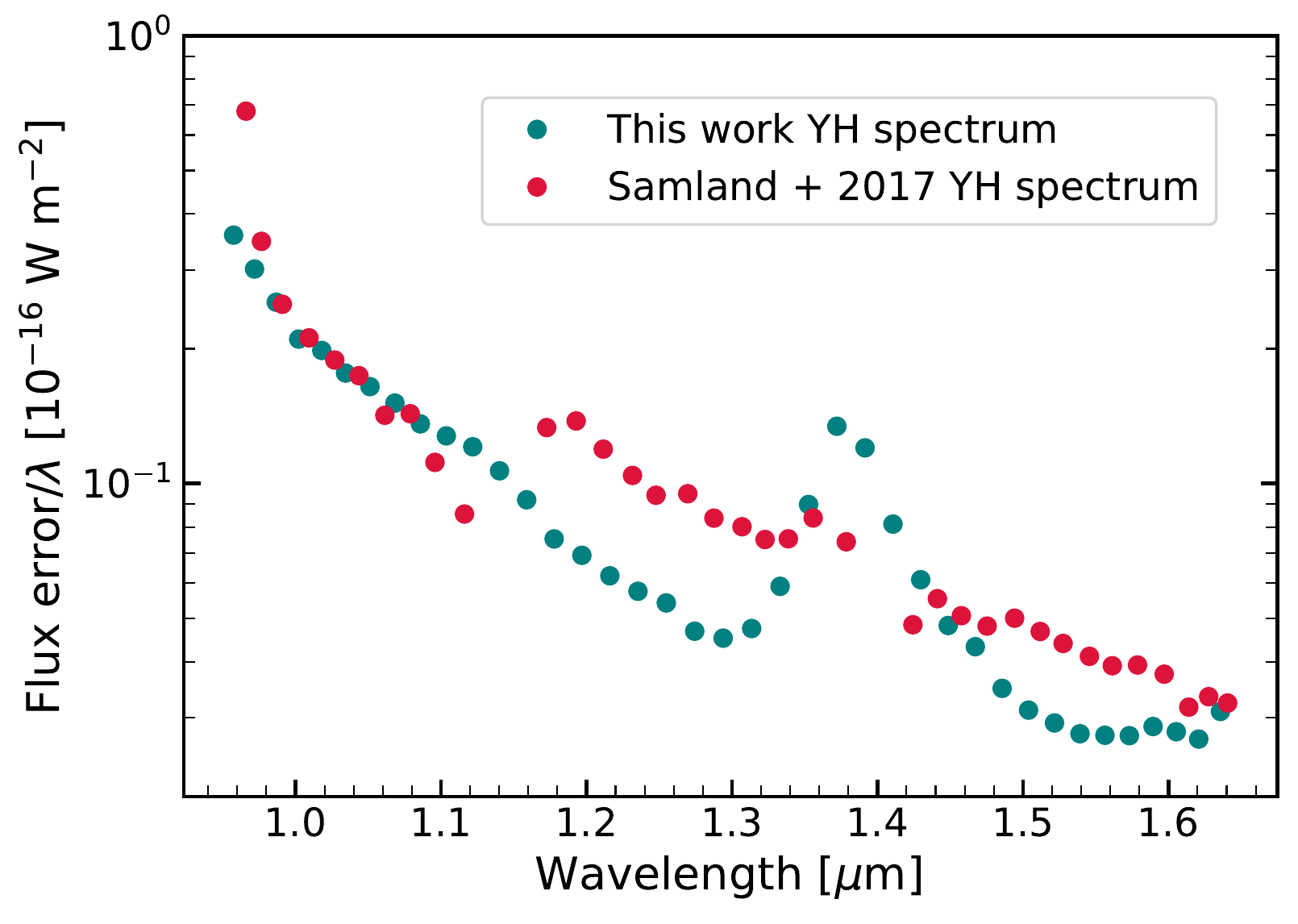}
 \caption{Flux uncertainty over wavelength as a function of wavelength for the $YH$ spectra of 51~Eri~b from this work compared to the one presented in \cite{Samland2017}.
              }
    \label{fig:comp}
\end{figure}

Figure~\ref{fig:spec} shows the spectrum of 51~Eri~b using our IFS data along with the $K1K2$ photometric points, overplotted for comparison is the IFS/$YH$ spectrum and the $K1K2$ photometry presented in \citetalias{Samland2017}, as well as the updated GPI $J$ and $H$ spectra and the $K1$ and $K2$ spectra from \cite{Rajan2017}. From the latter we see that the GPI $K1$ and $K2$ spectra is consistent with our derived $K1K2$ SPHERE photometry. It is also worth noticing that even with the revised version of the GPI $J$ and $H$ spectra, discrepancies with the SPHERE data still persist. Furthermore, we now see a slight discrepancy in the $H-$band spectrum as well, although mostly within error bars. As explained in detail in \citetalias{Samland2017}, the discrepancies can most likely be attributed to systematics. To further highlight the differences between the two IFS/$YH$ spectra, in Fig.~\ref{fig:comp} we present the relative error as a function of wavelength. For both figures, we removed the spectral channels that were not used in \citetalias{Samland2017} (1.14 and 1.41 $\mu$m), to allow for a better comparison. Overall, our data exhibits a lower relative error, except for the telluric H$_2$O absorption bands regions around 1.1 and 1.35 -- 1.4 $\mu$m (see Appendix \ref{sec:data_reduction_sphere_tellurics}), which results in a higher S/N ratio than the data in \citetalias{Samland2017}.

\begin{table*}
\caption[IRDIS photometry]{Photometry retrieved from the IRDIS data.} 
	\centering
	\begin{tabular}{ccccccccccc}
	\hline
    \hline
    	Filter & $\lambda$ & $\Delta \lambda$ & Flux & Contrast & Abs. magnitude \\
    	 & ($\mu$m) & ($\mu$m) & (Wm$^{-2}\mu$m$^{-1}$) &  &  \\
        \hline
		K1  & 2.110 & 0.102 & 4.418 $\times 10^{-17}$ $\pm$ 1.894 $\times 10^{-18}$ & 6.304 $\times 10^{-6}$ $\pm$ 2.703 $\times 10^{-7}$ & 15.11 $\pm$ 0.04 \\
		K2  & 2.251 & 0.109 & 5.149 $\times 10^{-18}$ $\pm$ 1.822 $\times 10^{-18}$ & 1.002 $\times 10^{-6}$ $\pm$ 3.546 $\times 10^{-7}$ & 17.11 $\pm$ 0.38 \\
        \hline
	\end{tabular}
  \label{tab:phot}
\end{table*}

\subsection{Detection limits}
\label{sec:det_lim}

\begin{figure}
    \centering
    \begin{subfigure}[]{1\textwidth}
        \includegraphics[width=0.47\linewidth]{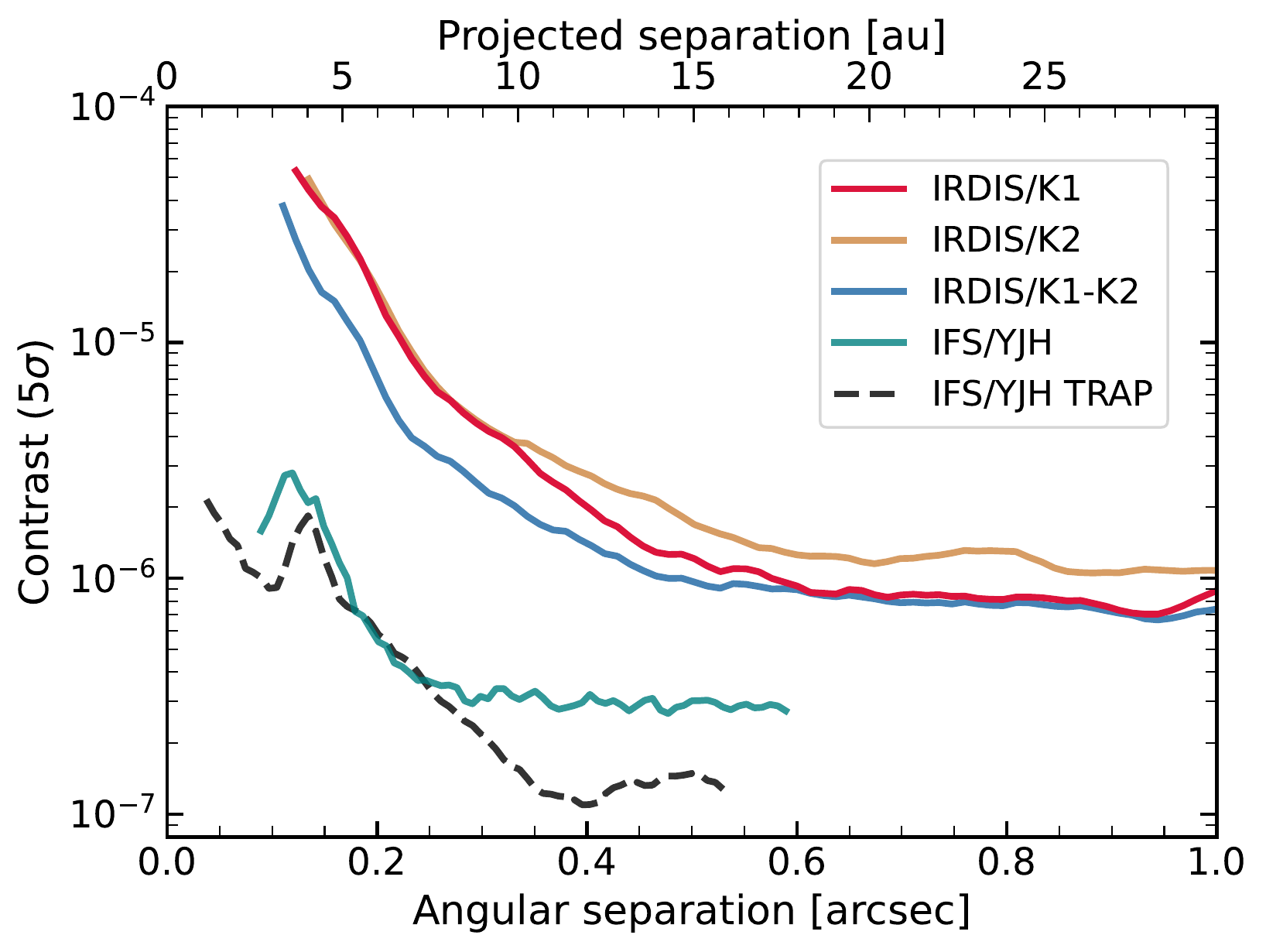}\\ 
    \end{subfigure}
    
    \begin{subfigure}[]{1\textwidth}
        \includegraphics[width=0.465\linewidth]{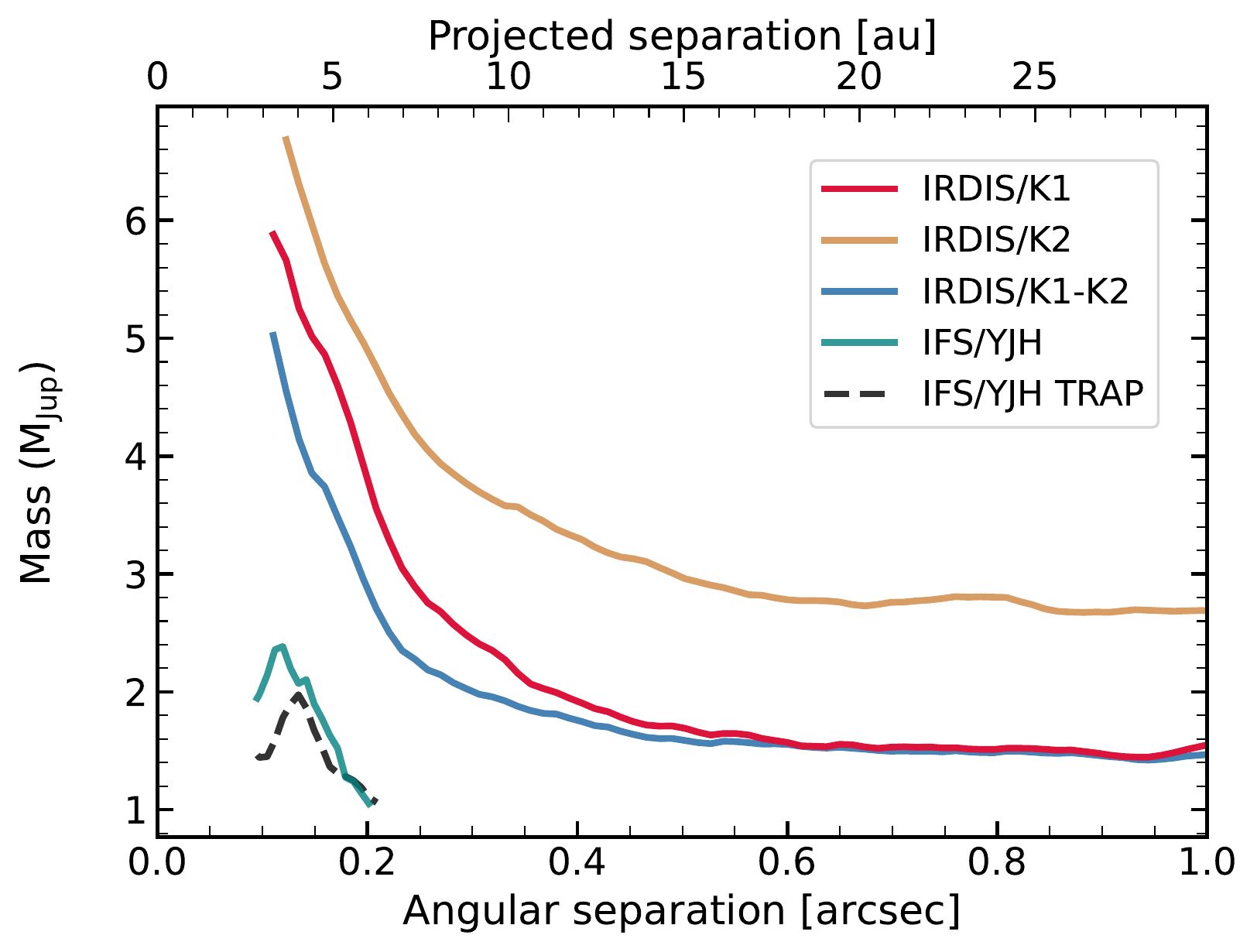}
    \end{subfigure}
    \caption[Contrast and mass limits]{5$\sigma$ detection limits. Planet contrast (\textit{top}) with respect to the star as a function of separation to the star, and planet mass detection limits (\textit{bottom}) as a function of the separation from the star for the SPHERE/IRDIS\_EXT data of 51~Eri~b. The mass limit for the IFS data is cut to the lowest mass computed by the model grid (1 M$_\mathrm{Jup}$) for both ANDROMEDA and TRAP.}
    \label{fig:contmass}
   \end{figure}

We used both ANDROMEDA and TRAP to derive 5$\sigma$ contrast curves for our IRDIS and IFS data. For the IRDIS/$K1$-$K2$ bands, the analysis setup was SDI+ANDROMEDA assuming no planet flux in $K2$. For the IFS-$YH$, we also used SDI+ANDROMEDA assuming a T5 spectral template for putative planets. To convert contrast to mass limits we used the evolutionary tracks of \cite{Baraffe2003} along with the atmosphere model of \cite{Baraffe2015}. For the star we used the 2MASS $JHK$ magnitudes \citep{Cutri2003}, the $L_P$ magnitude from \cite{Macintosh2015}, the newly calculated distance from \cite{Gaia2021}, and the latest age estimate from \cite{Lee2022}. The resulting detection limits are shown in Figure~\ref{fig:contmass}, we show the curves from ANDROMEDA and overplotted in dashed lines the curves obtained with TRAP for the IFS data.
The mass limit for the IFS data is cut to the lowest mass computed by the model grid for both pipelines. The $K1$ mass curve reaches smaller values than the $K2$ mass curve because the $K2$ filter matches a methane absorption band that strongly dims the flux of cold giant planets. Our detection limits are corrected for the coronagraphic transmission \citep{Boccaletti2018} and for small sample statistics \citep{Mawet2014}. From ANDROMEDA we get that the IFS data is sensitive to planets more massive than $\sim$1~M$_{\rm{Jup}}$ beyond $\sim$6~au, and 2~M$_{\rm{Jup}}$ beyond 4.5~au. While with TRAP we see an additional improvement in the sensitivity of planets down to 2~M$_{\rm{Jup}}$ at 3~au. We thus achieve a sensitivity about 2~M$_{\rm{Jup}}$ better than previous studies \citepalias[e.g.,][]{Samland2017}.

\section{Retrieval analysis}
\label{sec:retr}

We used the 1D radiative transfer code \texttt{petitRADTRANS}\footnote{\url{https://petitradtrans.readthedocs.io/}} \citep[\textit{pRT};][]{Molliere2019} in its scattering implementation \citep{Molliere2020} in combination with PyMultiNest\footnote{\url{https://johannesbuchner.github.io/PyMultiNest/}} \citep{Buchner2014} to derive the posterior distributions of the thermal structure, chemical composition, and cloud properties of 51\,Eri\,b. The code takes as an input the spectra of the planet along with prior distributions for the metallicity, $C/O$ ratio, log\,g, radius, a list of molecules to be included, quench pressure, and cloud parameters such as $f_{\rm{sed}}$, $K_{zz}$ and log~$\tau_{\rm{cloud}}$.

\subsection{Modelling setup}

\subsubsection{\texttt{petitRADTRANS} setup \label{sec:prt-setup}}

\begin{table}
\caption[Priors]{Prior values used for \texttt{petitRADTRANS} retrievals. \label{tab:prt-priors}} 
	\centering
	\begin{tabular}{ccccccccccc}
	\hline
    \hline
    	Model & Nominal & Both & Enforced clouds \\
        \hline
        [Fe/H] &  & $-1.5$-1.5 &  \\ 
        C/O &  & 0.1-1.6 & \\ 
        log $g$ &  & 2.0-5.5 & \\ 
        log $\tau_{\mathrm{cloud}}$ & $-7.0$-3.0 & & $-1.0$-3.0 \\ 
        f$_{\rm sed}$ &  & 0.0-10.0 & \\ 
        log K$_{zz}$ &  & 5.0-13.0 & \\
        Radius (R$_J$) &  & 0.9-2.0 & \\
        \hline
	\end{tabular}
\end{table}

The atmospheric retrieval modelling setup of \textit{pRT} is described in detail in \cite{Molliere2020}. In the following we describe the parameters used in our retrievals:

\paragraph{\textit{Retrieved parameters}}\mbox{}

The following parameters are of prime interest in the retrieval. We assign a flat prior to each - see Table~\ref{tab:prt-priors}.

\begin{description}
\item[Fe/H] The metallicity of the planetary atmosphere.

\item[$C/O$] The carbon-to-oxygen ratio prevalent in the planetary atmosphere.

\item[log~$g$] Logarithm of the surface gravity in units of centimeters per second-squared.

\item[$f_\mathrm{sed}$] The ratio of the mass averaged settling velocity of the cloud particles and the atmospheric mixing speed. Measures the efficiency of sedimentation in the atmosphere.

\item[log K$_{zz}$] Vertical eddy diffusion coefficient of the atmosphere.

\item[Radius] Of the planet's photosphere ($\tau=2/3$), in units of Jupiter radii.

\item[log~p$_\mathrm{quench}$] Logarithm of pressure at which carbon chemistry is quenched.

\item[$\sigma_\mathrm{lnorm}$] The geometric standard deviation in log-normal size distributions of condensates following \cite{Ackerman2001}.

\end{description}

\paragraph{\textit{Additional parameters}}\mbox{}

An important additional parameter is the effective temperature of the planet, $T_\mathrm{eff}$. This is not an explicit input (and thus retrievable) parameter of \textit{pRT}, but instead has to be calculated by generating a second spectrum for a given set of parameters that covers a wide spectral range in order to estimate the bolometric flux of the planet. Due to the required large wavelength coverage this can be quite time-consuming and is usually only carried out on a subset of the equal weighted posterior distribution.

In addition to these parameters, there are a number of internal ``nuisance'' parameters that also get prior ranges assigned. These are a set of connecting temperatures ($t_1$,$t_2$,$t_3$), an internal temperature $t_\mathrm{int}$, and the two parameters log~$\delta$ and $\alpha$ for the optical depth model $\tau = \delta P^\alpha$. These parameters are later used to determine the atmospheric pressure-temperature (\textit{P-T}) profile. This \textit{P-T} model is described in \cite{Molliere2020}. 

\paragraph{\textit{Clouds}}\mbox{}

From our first retrieval runs on the new data, we observed that the best-fit solutions tended to be non-cloudy ones. According to previously reported results, 51~Eri~b's photosphere is thought to be at least partially cloudy \citep[e.g.,][]{Samland2017, Rajan2017}. We decided to implement the parameter log~$\tau_\mathrm{cloud}$ to enforce clouds in the retrieval, which represents the logarithm of the cloud optical depth at the photospheric region of the clear atmosphere ($\tau = 1$). These are wavelength averaged optical depths, estimated over the wavelength range of the retrievals. By changing the range of the prior of log~$\tau_\mathrm{cloud}$ we were able to enforce clouds in the photospheric region.
In order to verify whether a cloud cover is actually present, we split our models into ``nominal'' and ``enforced clouds'' ones. Essentially, they are the same model, only differing in the prior range of log~$\tau_\mathrm{cloud}$, $[-7,3]$ for the ``nominal'', and $[-1,3]$ for ``enforced clouds'' (c.f.\,Sect.~\ref{tab:prt-priors}).  

For the cloud species, we use Na$_2$S and KCl, which according to \cite{Morley2012} are the most important species at the previously estimated effective temperature of 51~Eri~b (700-750 K).

\paragraph{\textit{Molecular species}}\mbox{}

The species contributing to the line opacities in our model are CO, H$_2$O, CH$_4$, NH$_3$, CO$_2$, H$_2$S, Na, K, PH$_3$, VO, TiO, and FeH. In addition, we include H$_2$ and He as species contributing to both Rayleigh scattering, and collision induced absorption. The species are retrieved under equilibrium chemistry assumptions and including quenching pressure.

\subsubsection{MultiNest setup}
To fit model spectra to the data by sampling the posterior probability, we used the nested sampling library PyMultiNest \citep{Buchner2014}, which is in turn based on MultiNest (\cite{Feroz2008,Feroz2009,Feroz2019}). Nested sampling (\cite{Skilling2004,Skilling2006}) is a powerful method, which in contrast to MCMC methods is better at exploring the parameter space and less prone to fall into a local minimum. 

Our derived model parameters are chosen to be the median of the marginalized, equal weighted posterior distribution, and the uncertainties quoted refer to the 16th and 84th percentile of said posteriors. Note that the parameters used to generate actual best-fit spectra are generally different from the medians mentioned above and corresponding to the highest log-likelihood. We use Importance Nested Sampling with flat priors (see Sect.~\ref{sec:prt-setup}), 4000 initial live points to sufficiently cover the parameter space, and to ensure a high sample acceptance fraction, we use MultiNest's constant efficiency mode and a sampling efficiency of 0.05.

\subsubsection{Input data}

\paragraph{\textit{Data Set}}\mbox{}

Our $YH$-spectrum, like all SPHERE/IFS spectra comprises 39 channels, all of which were fed into the retrieval process. In addition, we use the IRDIS $H$2/$H$3 photometric filters and broad-band $H$ from \citetalias{Samland2017}, and $K$1/$K$2 from this work. Also included are the $L_P$ and $M_S$ Keck/NIRC2 data points from \cite{Rajan2017}. We do not include GPI spectra in these retrievals due to the difficulty in quantifying systematic differences in the calibration of different instruments. A study combining data from SPHERE and $K$-band GRAVITY to expand the wavelength coverage using the retrieval approach, is reserved for future work (Samland et al., in prep.).

As described in \citetalias{Samland2017} and originally in \cite{Greco2016}, the spectral covariance of the residual speckle noise should be taken into account when computing the likelihood of a model matching data from an IFS-type instrument. We computed the correlation matrix $\Psi$ for our spectrum in the same way as \citetalias{Samland2017}. Consequently, in the log-likelihood computation in the retrieval code we used 

\begin{equation}
    -2 \log \mathcal{L} = (S-F)^{\mathrm{T}}C^{-1}(S-F),
\end{equation}

instead of the simple

\begin{equation}
    -2 \log \mathcal{L} = \Sigma_i ((S_i-F_i)/\sigma_i)^2.
\end{equation}

In both equations, $S$ represents the observed spectrum, and $F$ the model spectrum. 

See Appendix~\ref{sec:corr_mat} for more information.

The SPHERE $YH$ spectrum and the covariance matrix is available in Vizier/CDS.

\subsection{Results and comparison with older data}

\subsubsection{Retrieval results}

\begin{figure*}
\sidecaption
   \includegraphics[width=12cm]{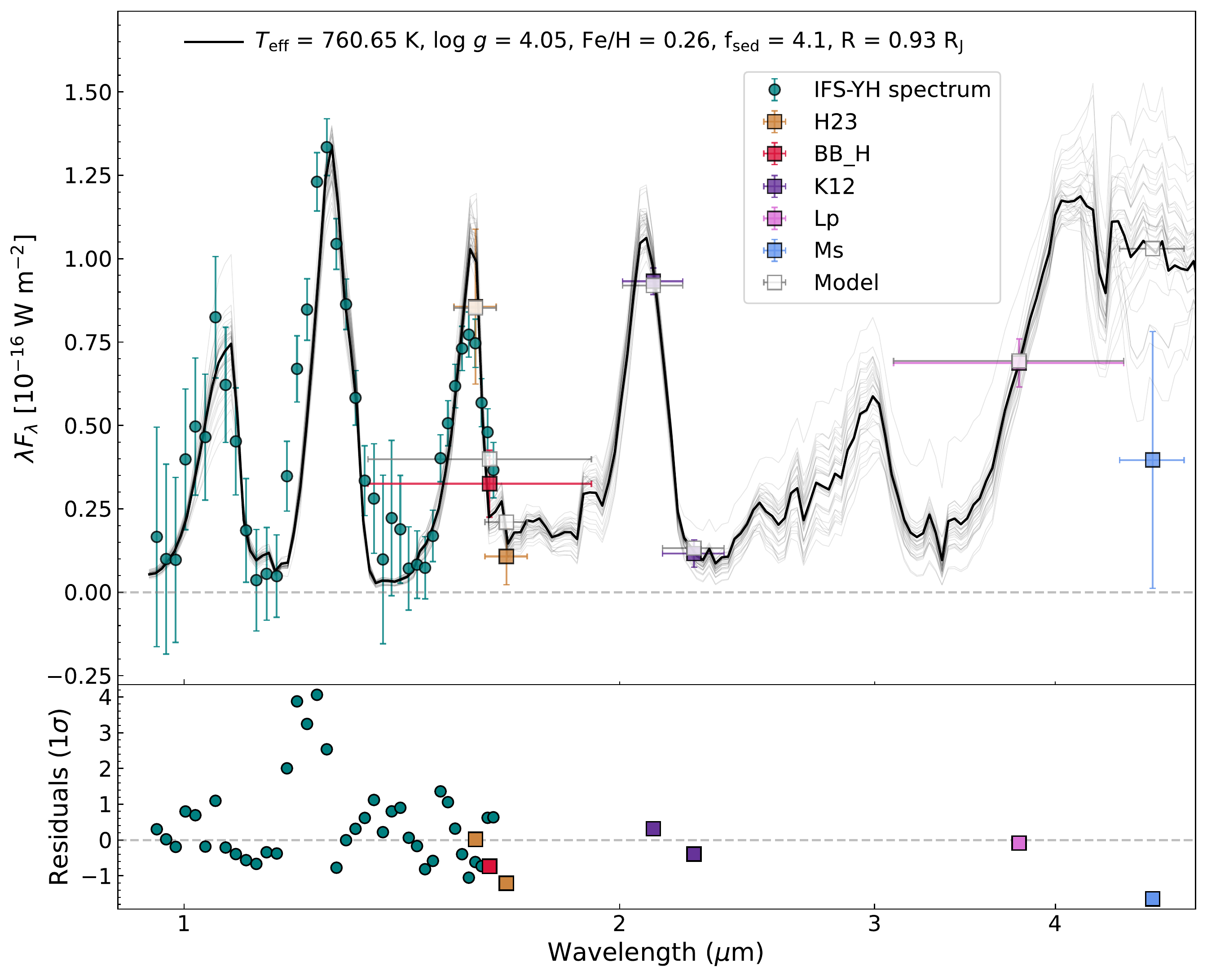}
    \caption{\texttt{petitRADTRANS} best-fit spectrum of 51~Eri~b for the ``nominal'' retrieval run (shown in black) on our new SPHERE spectro-photometric data (teal circles and purple squares) along with the photometric points included in \citetalias{Samland2017} and \cite{Rajan2017} (shown as squares). The photometric points describe the average flux in the respective filter, the \textit{x}-error bar represents the filter widths. 34 randomly drawn samples from the posterior probability distribution are shown in gray, to show the spread of model parameter combinations to fit the data. Residuals in multiples of 1 $\sigma$ uncertainties of the data for the best-fit model are shown below.}
    \label{fig:new_free}
\end{figure*}

\begin{figure*}
   \centering
    \includegraphics[width=0.9\textwidth]{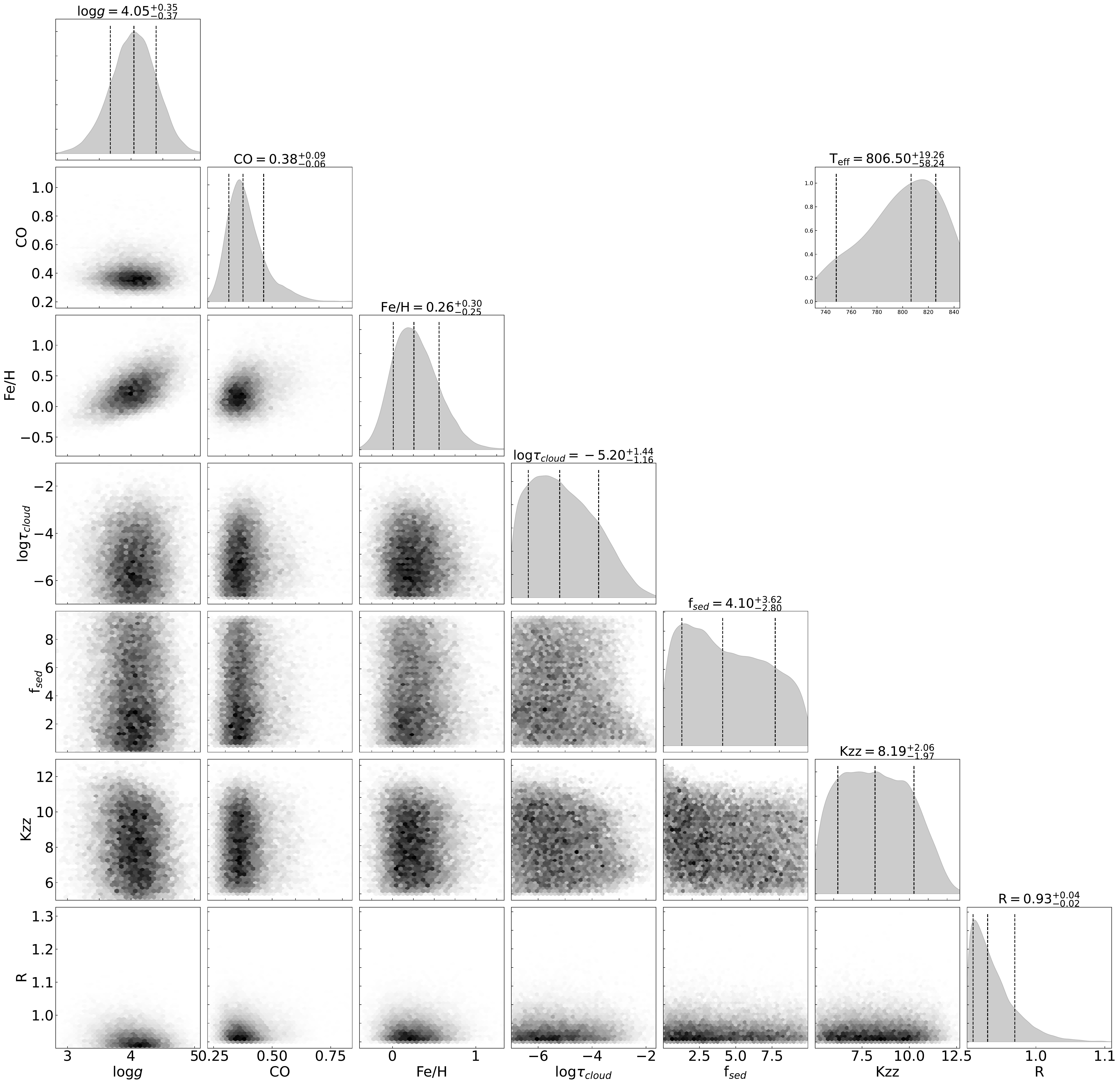}
   \caption{Corner plot of the posterior PDFs of the ``nominal'' retrieval run on the new data set.}
              \label{fig:free_corner}
\end{figure*}

In Fig.~\ref{fig:new_free} we show the best-fit spectra for the ``nominal'' model with the new data. Here the black line depicts the best fit spectrum, and the gray lines are randomly drawn spectra from the equally weighted posterior probability distribution. The circles represent the IFS spectrum of 51~Eri~b, while the squares depict the photometric points. Overall, the best-fit model is able to reproduce our $YH$ spectrum, as well as the $H2H3$, $K1K2$ and broad-band $H$ photometric points within the uncertainties. We note however, that the $J$-band flux from the best-fit model is lower than in our data by as much as $3-4\,\sigma$, which also occurs in the ``enforced clouds'' retrieval (see Fig.~\ref{fig:new_cloudy}). This discrepancy appears to be caused by systematics either in the opacities provided to the model, or, more likely, in the wavelength calibration strategy applied by the reduction pipeline that we are not able to remove. In either case, by using the correlation matrix these data points do not have a meaningful impact on the spectrum fit. Additionally, the $M_S$ photometric point is not fitted in either the ``nominal'' nor the ``enforced clouds'' retrievals. Given the large photometric uncertainties in the $M_S$ data, the best-fit photometry model lies within 2$\sigma$ of the data. As further discussed in Sect.~\ref{sec:discussion}, future mid-IR instruments might help to better constrain this photometric band.

Figure~\ref{fig:free_corner} shows the corresponding corner plot for the ``nominal'' model with the posterior probability density functions (PDFs) of selected parameters. The best-fit spectra along with the corner plots for the ``enforced clouds'' retrieval with our new data, are shown in Appendix~\ref{sec:repr_samland}.
The median of the posterior probability distribution, and the uncertainties representing a 1$\sigma$ uncertainty range for selected parameters (chosen due to their physical relevance) of our four sets of retrievals, are shown in Table~\ref{tab:results}. Here, the ``New'' in Cols.~2 and 3 refers to the retrieval results using our new data + the photometry in \citetalias{Samland2017} + \cite{Rajan2017}'s $M_S$ and $L_P$. 

{
\renewcommand{\arraystretch}{1.5}
\begin{table*}
\centering
\caption[Results]{Median and 1$\sigma$ uncertainties of the posterior probability distributions from the \texttt{petitRADTRANS} retrievals, using our new data and the data in \citetalias{Samland2017}. The last column shows the results of \citetalias{Samland2017} for comparison.} 
\begin{threeparttable}
	\begin{tabular}{cccccccc}
	\hline
	\hline
    Run & New nominal & New enforced clouds & SAM17 nominal & SAM17 enforced clouds & SAM17 \\ 
    \hline 
    [Fe/H] & 0.26 $\pm$ 0.30 & 0.29 $\pm$ 0.26 & -0.09 $\pm$ 0.20 & 0.03 $\pm$ 0.25 & 1.03$^{+0.10}_{-0.11}$ \\ 
    C/O & 0.38 $\pm$ 0.09 & 0.43 $\pm$ 0.07 & 0.80 $\pm$ 0.18 & 0.74 $\pm$ 0.16 & $-$ \\
    log $g$ & 4.05 $\pm$ 0.37 & 4.46 $\pm$ 0.38 & 4.53 $\pm$ 0.38 & 4.6 $\pm$ 0.4 & 4.26$^{+0.24}_{-0.25}$ \\ 
    log $\tau_{\rm clouds}$ & -5.20 $\pm$ 1.44 & -0.85 $\pm$ 0.16 & -4.7 $\pm$ 1.7 & -0.86 $\pm$ 0.17 & $-$ \\
    f$_{\rm sed}$ & 4.10 $\pm$ 3.62 & 0.25 $\pm$ 0.29 & 4.8 $\pm$ 3.5 & 0.32 $\pm$ 0.42 & 1.26$^{+0.36}_{-0.29}$ \\
    K$_{\rm zz}$ & 8.19 $\pm$ 2.06 & 7.58 $\pm$ 0.93 & 9.4 $\pm$ 2.9 & 7.9 $\pm$ 1.5 & 10$^{7.5}$ \\
    Radius (R$_{\rm Jup}$) & 0.93 $\pm$ 0.04 & 0.99 $\pm$ 0.09 & 1.17 $\pm$ 0.10 & 1.2 $\pm$ 0.1 & 1.11$^{+0.16}_{-0.13}$ \\
    T$_{\rm eff}$ (K) & 807 $\pm$ 45 & 744 $\pm$ 31 & 691 $\pm$ 22 & 634 $\pm$ 30 & 760$^{+21}_{-22}$ \\
    \hline
	\end{tabular}
	\begin{tablenotes}
	    \item {\footnotesize \textbf{Note.} The ``New'' in Cols.~2 and 3 refers to the retrieval results using our new data + the photometry in \citetalias{Samland2017} + \cite{Rajan2017}'s $M_S$ and $L_P$.}
	\end{tablenotes}
\end{threeparttable}
\label{tab:results}
\end{table*}
}

\subsubsection{Atmospheric retrieval on previous data}

In order to determine how far the use of a retrieval method alone impacts the outcome in terms of the derived atmospheric parameters with respect to \citetalias{Samland2017}'s grid of self-consistent models, we ran an additional set of \textit{pRT} retrievals (nominal and enforcing clouds) on the original data used by \citetalias{Samland2017}. These attempts are also useful to understand to what extent the outcomes are a result of the better quality of the SPHERE 2017 data. We included all the data cited in \citeauthor{Samland2017} work, i.e. the $YJ$ and the $YH$ spectra, and the $H2H3$, $K1K2$, $L_P$, and broad-band $H$ photometric points, all resulting from 2015 and 2016 observations, as well as the GPI spectrum published in the discovery paper \citep{Macintosh2015}. Note that \citetalias{Samland2017} used an SDI+ANDROMEDA data reduction resulting in the reference channels of their spectra not being usable in the model comparison. In addition, they masked out a number of channels due to low signal-to-noise ratios. We took these considerations into account in our reproduction attempts. We do not include the revised version of the $J$ and $H$ spectra, nor the $K1K2$ spectra from \cite{Rajan2017} to be consistent in our attempt to reproduce \citetalias{Samland2017}'s results. The results of the nominal and enforced clouds retrievals are shown in Appendix~\ref{sec:repr_samland}.

When comparing the parameter values for the ``SAM17 nominal'' retrieval in Table~\ref{tab:results} to the ones in \citetalias{Samland2017}'s Table~5 (top row,``PTC-C'', shown in the last row of Table~\ref{tab:results}), we notice that we find a significantly lower metallicity, and very little indication for the presence of clouds. Other parameters differ as well, but are mostly within the 16/84th percentile limits. We discuss this issue, and in particular our attempt to determine whether the object is cloudy or not, in depth in Sect.~\ref{sec:retrieval-scm}.

\section{Discussion}
\label{sec:discussion}

\subsection{Parameters of 51~Eri~b}

In the following we present a discussion of certain parameters of interest of 51~Eri~b.

\subsubsection{C/O ratio}

The atmospheric carbon-to-oxygen ratio has been linked to the formation scenario of exoplanets \citep{Oberg2011}. The different condensation temperatures of water (H$_2$O), carbon oxide (CO) and carbon dioxide (CO$_2$) locate their ``ice lines'' in different parts of the protoplanetary disk, which results in different values of C/O through the disk. A difference in the C/O ratio of a gaseous giant planet compared to its host star's C/O ratio can provide information about the planet's formation, depending on whether their C/O ratio is super- or sub-stellar. See, however, \cite{Molliere2022} on how challenging it is to go from C/O to formation, in practice.

We find the C/O ratio of 51~Eri~b to be consistent along retrieval runs for the same data set (C/O $\sim$ 0.4 $\pm$ 0.08 for the new data and C/O $\sim$ 0.8 $\pm$ 0.17 for the data in \citetalias{Samland2017}), respectively. Compared to the solar C/O ratio assumed by \citetalias{Samland2017} (i.e. C/O = 0.55), all our retrieved values within error bars differ by $\sim$0.1. However, no definitive conclusions can be drawn since the value for the C/O ratio of the star remains undetermined. The lower flux from the $M_S$ photometric point hints to the presence of carbon monoxide (CO), however, further observations would be needed to use it to constrain the C/O ratio of 51~Eri~b.

\subsubsection{[Fe/H]}

In our retrievals using the data in \citetalias{Samland2017} we obtain a metallicity [Fe/H] in the range of -0.09 to 0.30, including uncertainties, which differs from the results of \citetalias{Samland2017} who find [Fe/H] = 1.0$\pm$0.1. However, when comparing with retrieval results for two benchmark brown dwarfs, the authors find that they tend to fall in the lower end of the estimated metallicity range for the host star. In our case, we observe a similar behaviour, even for the new data we obtain [Fe/H] $\sim$ 0.26 which is slightly super-solar. As discussed at length in \citetalias{Samland2017}, the derived metallicity strongly depends on the $K$-band flux, and indeed our retrieval tends to slightly underfit the $K1$-flux, whereas it was overfitted in \citetalias{Samland2017}. When artificially enhancing the importance of the $K1$-flux point by lowering its uncertainty by a factor 10, the resulting fit for \citetalias{Samland2017}'s data shows an increased metallicity ([Fe/H] = $0.31^{+0.12}_{-0.13}$), and a higher log$\tau_{\mathrm{cloud}} = -4.2^{+1.28}_{-1.18}$. These values are closer to \citetalias{Samland2017}'s, but still do not agree. The initial and the remaining difference indicate that parameters derived from retrievals can differ significantly from self-consistent models.

\subsubsection{Clouds and log $\tau_\mathrm{clouds}$}
\label{sec:p-t_disc}

Our initial retrievals always resulted in the best-fit models tending towards non-cloudy solutions. Based on previous results, which suggested 51~Eri~b's atmosphere to be at least partially cloudy, we included an additional prior to enforce clouds and check the robustness of the retrieval: log $\tau_{\rm{clouds}}$. However, even for the ``enforced clouds'' retrievals, the value for log $\tau_{\rm{clouds}}$ always tended towards the lower limit of the prior (i.e. towards cloud-free solutions; see Table~\ref{tab:results}). Both $f_{\rm{sed}}$ and K$_{\rm{zz}}$ describe the cloud properties \citep{Ackerman2001}.
Our best-fit values for $f_{\rm{sed}}$ in the ``enforced clouds'' retrievals are in agreement with previously reported $f_{\rm{sed}}$ values for brown dwarfs in a similar temperature range as 51~Eri~b \citep[e.g., GJ 758 B, GJ 504 b;][, respectively]{Vigan2016, Skemer2016}. We also note that our values are within the ranges of $f_{\rm{sed}}$ found by \citetalias{Samland2017}. A higher $f_{\rm{sed}}$ corresponds to vertically thinner clouds with larger particle sizes. Our derived values for K$_{\rm{zz}}$ are within the assumed values in both \cite{Macintosh2015} and \citetalias{Samland2017}.

\begin{figure*}
 \centering
 \includegraphics[width=0.9\textwidth]{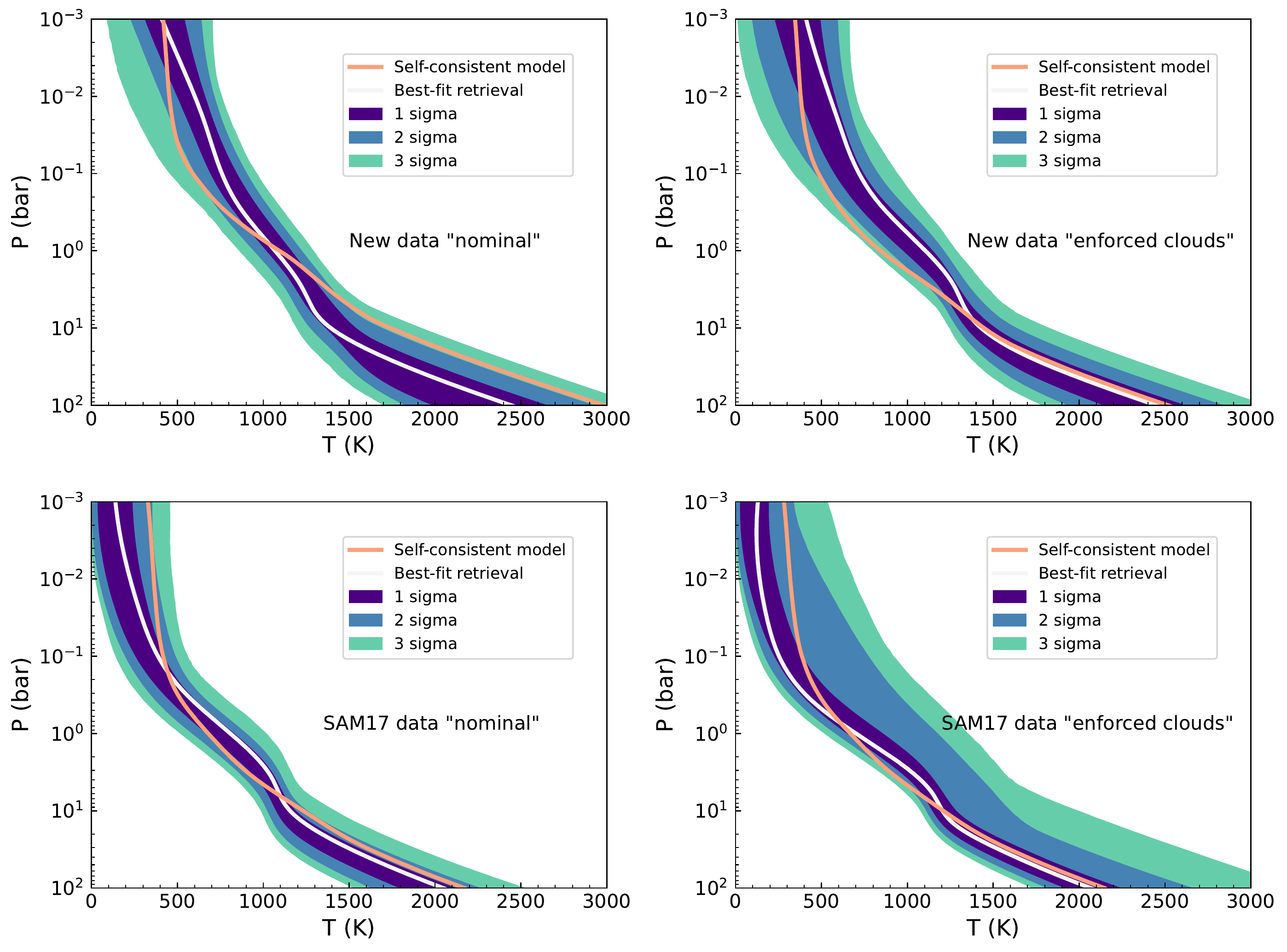}
 \caption{Retrieved pressure-temperature profiles in white, with confidence intervals (1-, 2- and 3$\sigma$) for our four different retrievals shown in Table \ref{tab:results}. Overplotted are the resulting self-consistent \textit{P-T} structures in pink.}
    \label{fig:pt}
\end{figure*}

In Fig.~\ref{fig:pt} we show the \textit{P-T} profiles resulting from the retrieval runs along with the 1-, 2- and 3$\sigma$ confidence intervals for our four different cases. Overplotted are the corresponding self-consistent \textit{P-T} profiles obtained from \texttt{petitCODE} when feeding in the best-fit parameters of the retrievals. We note that the self-consistent \textit{P-T} profiles are less isothermal than the retrieval ones in all cases, following the characteristic atmospheric temperature gradient for models in radiative-convective equilibrium. In this scenario, the only way to reproduce the observed NIR low fluxes is to add clouds, which contradicts our results with the retrieval models. This discrepancy has been observed in other studies (see Sect.~\ref{sec:retrieval-scm} for an in-depth discussion), and it is yet to be resolved in order to draw conclusions about the cloudiness of exoplanet atmospheres. 

\subsubsection{Radius and surface gravity}

The values we obtain for the radius and surface gravity are in agreement with previous results for the data in \citetalias{Samland2017} within uncertainties. As seen in Fig.~\ref{fig:free_corner} and Fig.~\ref{fig:cloudy_corner}, the retrieval finds the best-fitting models towards the lower $R_p$ prior boundary for the new data, while this is not the case for the retrievals using the data in \citetalias{Samland2017} (Fig.~\ref{fig:repr_samland_free_corner} and Fig.~\ref{fig:repr_samland_cloudy_corner}). We attribute these variations to the different input data. Nevertheless, all values are consistent with the radius of Jupiter within uncertainties, which according to planetary evolution models are slightly smaller than expected for the possible age of the system \citep[i.e. 10 - 20 Myrs, $\sim 1 -1.2\,\mathrm{R_{Jup}}$;][]{Mordasini2012}.

\subsubsection{Mass}

To derive an estimate of the mass of 51~Eri~b we used the posterior distribution for the surface gravity and radius of our ``nominal'' model, and the relation $M = g/g_{\rm{Jup}} \cdot (R/\mathrm{R_{Jup}})^2$, where $g_{\rm{Jup}} = 24.79$ m s$^{-2}$, and R$_{\rm{Jup}} = 6.99 \times 10^7$ m are the surface gravity and the volumetric mean radius of Jupiter, respectively. We obtain a mass of $M = 3.9 \pm 0.4 \mathrm{M_{Jup}}$. Additionally, we used the photometric values for the $K1K2$ bands along with the evolutionary models for extrasolar giant planets from \cite{Baraffe2003}. We used both estimates of the age of the system and we obtain a value of $M = 2.4 M_{\rm{Jup}}$ at 10 Myr, and $M = 2.6 \mathrm{M_{Jup}}$ at 20 Myr. All mass values of the planet are smaller than the value found by \citetalias{Samland2017} ($M = 9.1^{+4.9}_{-3.3} \mathrm{M_{Jup}}$), however, the formation scenario can strongly impact the mass (and the surface gravity) of the planet, and small masses are possible within the ``hot-'' and ``warm-start'' scenarios.

\subsection{Atmospheric retrievals vs self-consistent models\label{sec:retrieval-scm}}

Atmospheric retrievals are now a commonly used tool for fitting the spectra of exoplanets \citep[for a recent review, see][]{madhusudhan2019}. The general idea of retrievals is that an atmospheric forward model calculates planetary spectra based on a parameterized atmospheric structure, which is then compared to the data. This parameterization is key, because finding the atmospheric temperature, composition, and cloud structure in a physically self-consistent way is a numerically expensive step. Self-consistent models typically assume that the atmosphere is in radiative-convective equilibrium, and couple this assumption to a model solving for the atmosphere's chemical composition and cloud structure in an iterative fashion \citep[see e.g.,][for a review]{hubeny2017}. Furthermore, many processes, especially those connected to clouds, are not sufficiently well understood. If the underlying assumptions in the self-consistent model are incorrect, this may lead to unfounded conclusions about the atmospheric model's properties and parameters. At the same time, self-consistent models allow us to compare our complete physical understanding against what the data shows. 
Conversely, retrievals attempt to constrain the atmospheric structure mostly from the data alone (if uninformative priors are chosen), circumventing these issues. This requires data of high enough S/N and spectral coverage, however.

For cloudy directly imaged planets retrievals have proven challenging thus far. This is connected to a degeneracy, or at least a correlation: a cloud generally reddens the spectrum by hiding the deep hot regions of the atmosphere from view. If the cloud was not present, too much flux would escape from the atmosphere, especially in the opacity windows in the $Y$, $J$, and $H$ bands. Moreover, an atmospheric model in radiative-convective equilibrium generally results in a large atmospheric temperature gradient, such that the only way to reproduce the red spectral energy distributions (SEDs) of cloudy planets and brown dwarfs is to add clouds. In \citet{tremblinamundsen2015,tremblinamundsen2016,tremblinchabrier2017,tremblinpadioleau2019} atmospheric instabilities that decrease the atmospheric temperature gradient when compared to the equilibrium solution have been suggested to mimic the effect of clouds in the NIR. This can be easily understood: if the deep atmosphere is less hot, no clouds are required to lead to a reduced flux in the $YJH$-bands. For retrievals with a flexible atmospheric temperature and cloud parameterization this may thus result in atmospheric structures that are cloud-free and too isothermal when compared to classical self-consistent models. Due to the ease with which red exoplanet spectra can be fit with too-isothermal temperature profiles, it is not surprising that a retrieval can fall into this ``trap'': different temperature structure realizations are explored during a retrieval, and if the cloud model is not the ``perfect'' one, and leads to a slightly worse fit, there is no reason for the retrieval to add a cloud of appreciable opacity to the atmosphere.

This cloud-free retrieval problem appears to be emerging in recent studies \citep{Molliere2020,zhangsnellen2021b,kammererlacour2021}, and we also see it occurring here. Options to enforce a more cloudy solution may include making the atmospheric temperature parameterization less flexible, such that always a strong temperature gradient is present, which then needs to be corrected for by adding a cloud (which was identified as a workaround in \citealt{kammererlacour2021} when using the atmospheric retrieval code \texttt{ATMO}; \citet{tremblinamundsen2015, tremblinamundsen2016}). The danger is that such simple temperature profiles may not be complex enough to capture the atmospheric temperature structure even for a cloud-free planet, leading to potentially biased results for the atmospheric properties. Enforcing a minimum cloudiness in the atmosphere via a cloud optical depth prior, as we attempted to do here, may be another option, but we remind the reader that the retrievals still attempted to be as cloud-free and isothermal as possibly allowed. A truly satisfactory method for approaching this cloud-P-T correlation has yet to be found.

It is conceivable that these atmospheres are truly mostly cloud-free, and that this is the reason for the retrievals to tend towards these solutions, as also argued for in the Tremblin et al. papers. However, even synthetic cloudy spectra are retrieved to be cloud-free if the cloud model is modified between the synthetic observation and retrieval step \citep{Molliere2020}, such that the absence of clouds in the retrievals cannot be regarded as a proof of the absence of clouds in the atmospheres of real planets. A promising avenue is presented in \cite{burninghamfaherty2021}. The authors found that adding mid-IR data tracing silicate cloud absorption features at 10~$\mu$m will lead to definitely cloudy solutions, as well as temperature structures which are less isothermal compared to the retrievals that neglected the mid-IR data \citep{Burningham2017}. The James Webb Space Telescope's (JWST) observations of cloudy exoplanets and brown dwarfs with its mid-IR instrument MIRI \citep{Wright2004}, may thus hold great potential to resolve, at least partially, the cloud-temperature gradient correlation. A remaining challenge is that the 10~$\mu$m region probes lower pressures than the NIR ($YJH$ bands), thus probing the silicate feature could merely help to constrain the \textit{P-T} profile in the upper atmosphere. Indeed, \cite{burninghamfaherty2021} need to include a second deep cloud to produce the NIR reddening. However, we note that the above could be a general solution for planets but could not be applied to 51~Eri~b, since the planet is too cold to have silicate clouds, and too close to the host star to be observed with JWST/MIRI. A possible solution for cool, closer-in giant planets like 51~Eri~b, could be to use the Mid-Infrared ELT Imager and Spectrograph \citep[METIS, ][]{Quanz2015} to expand the walength coverage (up to 13~$\mu$m) and search for the silicate feature. Finally, we advocate that results from atmospheric retrievals should never be discussed in isolation, but instead along with the results from self-consistent models. This can be done in different steps:

\begin{enumerate}
    \item Compare to self-consistent atmospheric structures, obtained from using the retrieval's best-fit parameters for the atmospheric composition, gravity, effective temperature, and cloud parameters. This includes comparing pressure-temperature structures, as described in Sect.~\ref{sec:p-t_disc}, but also abundance structures with the expected values from chemical equilibrium \citep[e.g.,][]{Line2015, Line2017, Gandhi2018, Zalesky2019}. We note that in our case, chemical equilibrium is already imposed within the \textit{pRT} retrievals.
    \item Compare the retrieved cloud location and abundances with the expectations from physics and chemistry (e.g., compare the intersection of the saturation vapor pressure curve and the \textit{P-T} profile given the expected cloud base, with the retrieved location of the cloud; see Fig. 3 of \citealt{Burningham2021}). Once more, in our case, this is already included in the \textit{pRT} retrievals.
    \item Analyse robustness of the results when changing the model setup. Regarding the cloud model (i): our model is very strongly physically and chemically motivated, but another option would be a power law opacity implementation to mimic the clouds \citep[see Cloud Model 2 in][]{Molliere2020}; or various other  prescriptions \citep[see for example:][]{Burningham2017, Barstow2020}. Regarding the \textit{P-T} parameterization \citep[see for example, the discussion in][]{Molliere2020} (ii): the \textit{P-T} parameterization could also effectively be changed by imposing certain priors during the retrieval, for example by limiting the second derivative of the temperature structure such that solutions that would give a smooth \textit{P-T} profile are favored, as proposed by \cite{Line2015}, their Sect.~2.4.2. Our current retrievals do not have this implemented, but it can be added in our future work. In general, we note that robustness of retrieval results when changing the parameterization does not necessarily indicate correctness.
\end{enumerate}


\section{Summary and conclusions}
\label{sec:summary}

In this work we present VLT/SPHERE spectro-photometric observations of 51~Eridani b. The new $YH$ spectrum and $K1K2$ observations show improved S/N compared to previously reported data, allowing us to revise the published flux measurements. We used the radiative transfer code \texttt{petitRADTRANS}, which utilizes a retrieval approach to fit the atmospheric parameters. In addition, we attempted to reproduce previous results (obtained with self-consistent models) using this approach and compared the outcomes of retrievals to self-consistent models. 
Our results can be summarized as follows:

\begin{description}
    \item[$\bullet$ ] We extracted the spectrum of 51~Eri~b using the ANDROMEDA algorithm (Fig.~\ref{fig:spec}). We obtained new photometric measurements for the $K1K2$ filters ($M_{K1} = 15.11 \pm 0.04$ mag, $M_{K2} = 17.11 \pm 0.38$ mag; Table~\ref{tab:phot}).
    \item[$\bullet$ ] The detection limits derived from our data show an increased sensitivity and rule out the presence of planets more massive than 2~M$_{\rm{Jup}}$ at 3~au, and 1~M$_{\rm{Jup}}$ beyond 4.5 au (Fig.~\ref{fig:contmass}).
    \item[$\bullet$ ] Our initial retrieval runs tended towards clear atmospheres, to verify the robustness of these results we introduced an additional fit parameter (log $\tau_{\rm clouds}$) to enforce clouds. We report the results of four different cases in Table~\ref{tab:results}: a ``nominal'' and an ``enforced clouds'' version for our new data + the photometry in \citetalias{Samland2017} + \cite{Rajan2017}'s $M_S$ and $L_P$; and for the same data used in \citetalias{Samland2017}.
    \item[$\bullet$ ] We are able to obtain a good fit to the observations with \textit{pRT} (e.g. Fig.~\ref{fig:new_free}), with the exception of the $M_S$ photometric point, which can be explained by the large uncertainty of the data. Further mid-IR observations in this band could improve the fit and help constrain the C/O ratio of the planet. We observe, that even the ``enforced clouds'' retrieval runs tend towards non-cloudiness (log $\tau_\mathrm{clouds} = -0.85 \pm 0.16$), which differs from previous results obtained using self-consistent models. This discrepancy may be due to the larger and more flexible parameter space that can be explored with retrievals as opposed to self-consistent models. In particular, the isothermal \textit{P-T} profiles may imitate the effect of clouds.
    \item[$\bullet$ ] Overall, our results (C/O $= 0.38\pm0.09$, [Fe/H] $= 0.26\pm$0.30 dex, T$_{\rm{eff}} = 807\pm$45 K and log $g = 4.05\pm0.37$) are in agreement with previously reported parameters of the planet. One of the major disagreements is the metallicity, which we find to be close to solar with the new data. Once more, this can be explained by the different methods that atmospheric retrievals and self-consistent models use to fit the data. We estimate the mass of the planet to be between 2 and 4 M$_{\rm{Jup}}$, which is consistent with both ``hot-'' and ``warm-start'' formation scenarios.
    \item[$\bullet$ ] As an additional test, we used the best-fit parameters from the retrievals to obtain the pressure-temperature structure using a self-consistent model (Fig.~\ref{fig:pt}). The results show a larger temperature gradient for the self-consistent models, suggestive of the $T$-gradient-cloud correlation playing a role.
\end{description}

Our results highlight the challenges that are still to overcome when modelling exoplanet atmospheres, as well as the importance of observations at longer wavelengths to determine the presence or absence of clouds. Observations with future instruments that allow the study of additional cloud absorption features such as ELT/METIS, would be required to provide a final conclusion on the cloud-temperature gradient degeneracy.

\begin{acknowledgements}
      SPHERE is an instrument designed and built by a consortium consisting of IPAG (Grenoble, France), MPIA (Heidelberg, Germany), LAM (Marseille, France), LESIA (Paris, France), Laboratoire Lagrange (Nice, France), INAF - Osservatorio di Padova (Italy), Observatoire de Gen\`{e}ve (Switzerland), ETH Z\"{u}rich (Switzerland), NOVA (Netherlands), ON ERA (France) and ASTRON (Netherlands) in collaboration with ESO. SPHERE was funded by ESO, with additional contributions from CNRS (France), MPIA (Germany), INAF (Italy), FINES (Switzerland) and NOVA (Netherlands). SPHERE also received funding from the European Commission Sixth and Seventh Framework Programmes as part of the Optical Infrared Coordination Net- work for Astronomy (OPTICON) under grant number RII3-Ct-2004-001566 for FP6 (2004-2008), grant number 226604 for FP7 (2009-2012) and grant number 312430 for FP7 (2013-2016). This work has made use of the SPHERE Data Centre, jointly operated by OSUG/IPAG (Grenoble), PYTHEAS/LAM/CeSAM (Marseille), OCA/Lagrange (Nice), Observatoire de Paris/LESIA (Paris), and Observatoire de Lyon (OSUL/CRAL). A.L.M. acknowledges financial support from the Agence Nationale de la Recherche, the European Research Council under the European Union’s Horizon 2020 research and innovation program (Grant Agreement No. 819155), and the F.R.S.-FNRS.
\end{acknowledgements}

%
\bibliographystyle{aa} 
\bibliography{biblio} 
%

\begin{appendix} 

\section{Wavelength calibration and spectral differential imaging\label{sec:data_reduction_sphere}}

SPHERE/IFS data reduction pipelines remove instrumental signatures, calibrate and compute wavelength solutions, extract spectra of individual lenslets from the 2D detector, and re-assemble the data into a 3D data cube with one spectral and two spatial dimensions. The wavelength calibration relies on a range of monochromatic lasers projected on the detector by the calibration unit. 

The default of ESO's EsoRex pipeline \citep{Freudling2013} and the Data Reduction Handling software used by the SPHERE consortium \citep{Pavlov2008} is to determine the wavelength solution by fitting a 2nd order polynomial to the spectral calibration data. In addition to using spectral lines for the absolute wavelength calibration, the vlt-sphere Python package \citep{Vigan2020}\footnote{https://github.com/avigan/SPHERE} aims at a more refined calibration of the dispersion solution by tracing the radial separation between diagonally opposite satellite spots for each spectral plane in the 3D data cube.

\begin{figure}
       \begin{center}
        \includegraphics[width=0.49\textwidth]{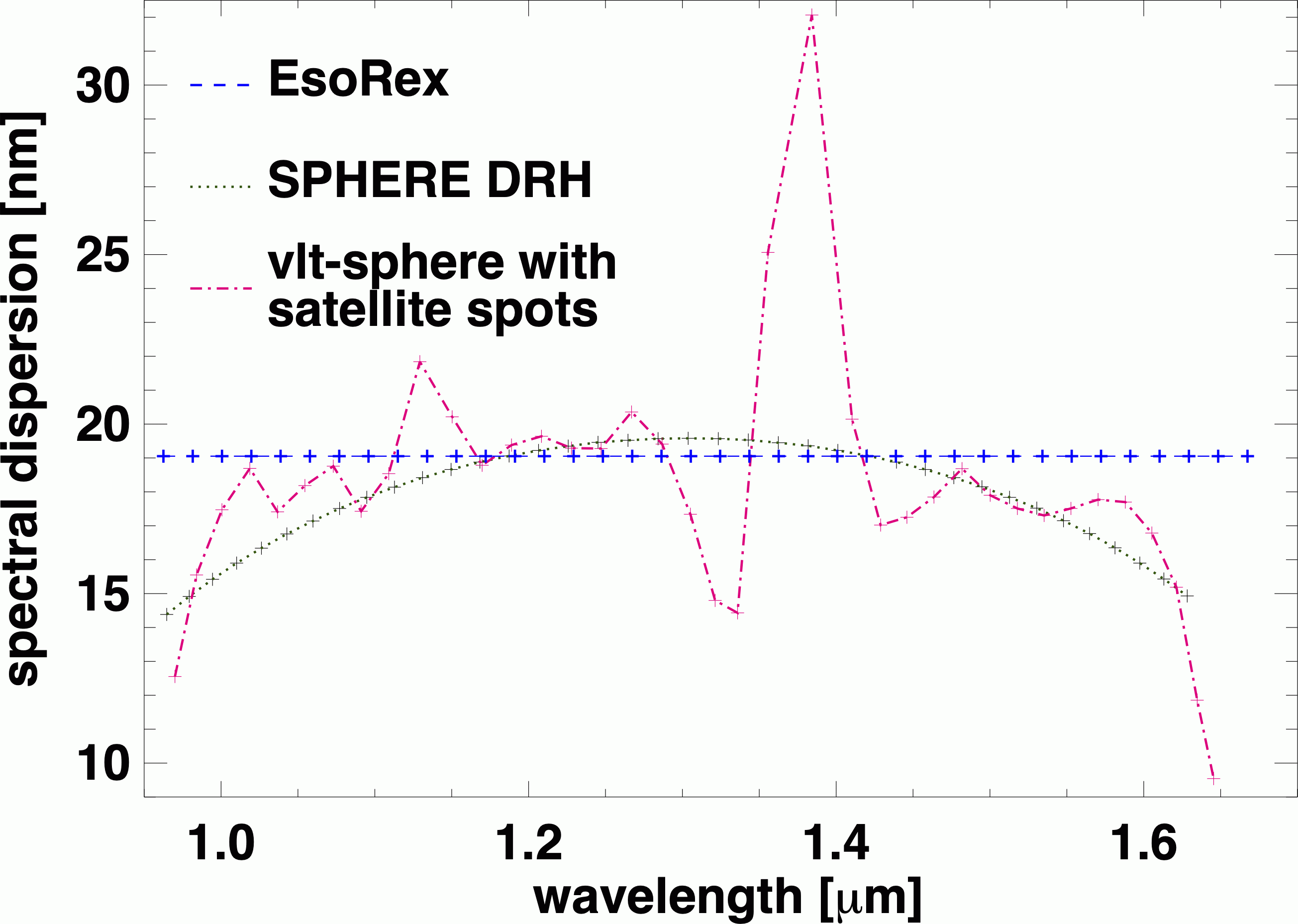}
        \includegraphics[width=0.46\textwidth]{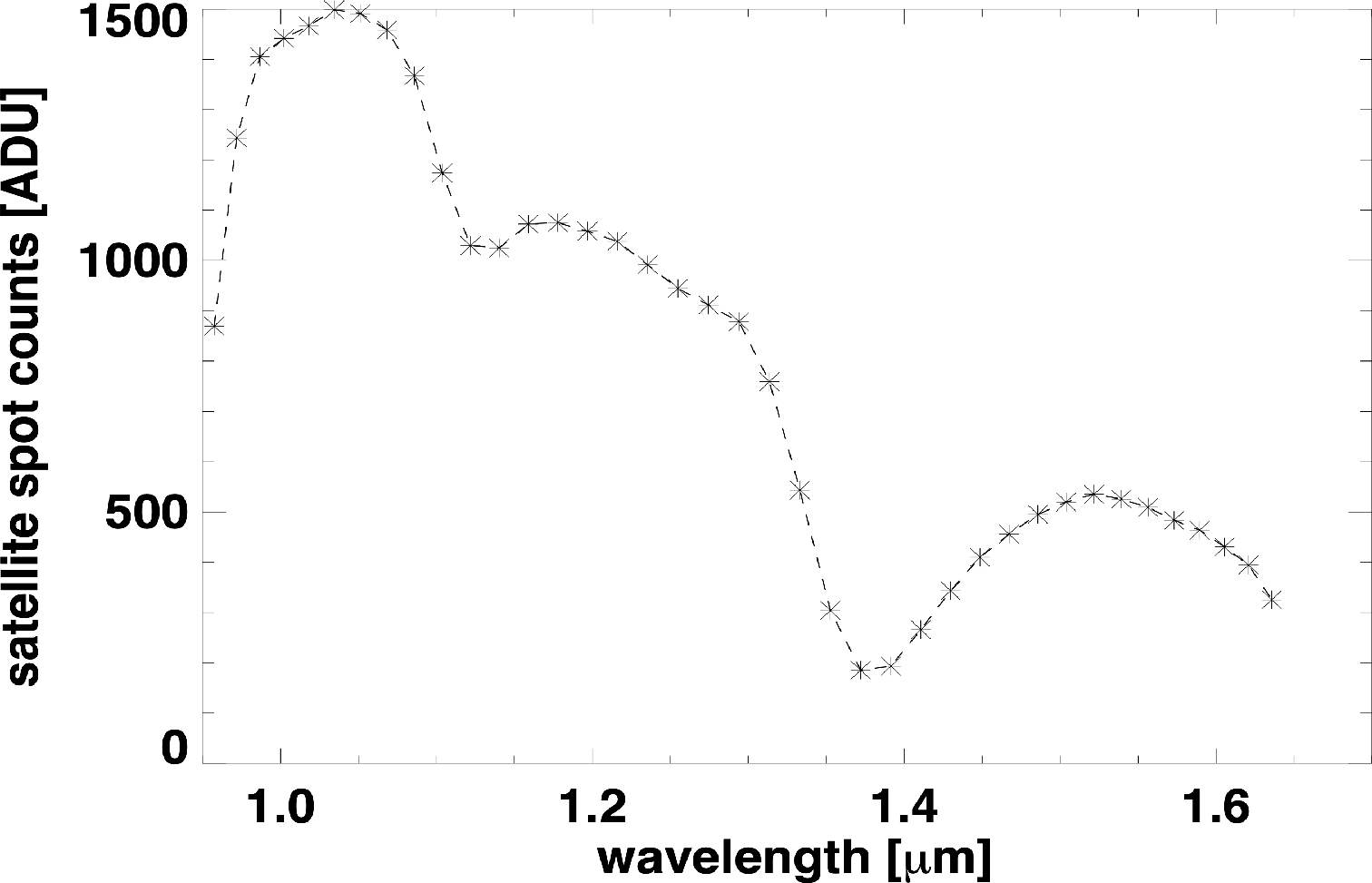}
    \end{center}
      \caption{Top: SPHERE/IFS dispersion solutions of the reduced 3D data cubes resulting from the standard ESO pipeline EsoRex, the Data Reduction and Handling (DRH) of the SPHERE consortium, and the vlt-sphere Python package tracing the separations of satellite spots. Bottom: the satellite spots show strong spectral gradients in the count rates at the edges of telluric absorption bands.}
    \label{fig:dispersion}
   \end{figure}

Figure~\ref{fig:dispersion} (top) visualizes the respective dispersion solutions inherent to three different IFS pipelines. Data cubes reduced by EsoRex version 0.42.0 include the median of the shortest wavelengths, and the median dispersion of a linear fit to the wavelength solution as keywords in the FITS header. SPHERE DRH and vlt-sphere (version 1.4.3, with a wavelength calibration issue fixed) provide a separate FITS file with wavelengths corresponding to each spectral plane of the 3D data cube. In the case of SPHERE DRH, this is based on the 2nd order polynomial fit. 

As the choice of dispersion solution determines the spectral band-width of individual spectral channels, it also influences the recovered spectral energy distribution of the detected astrophysical sources. This has to be considered, e.g., when applying retrieval techniques to the observational spectra. For Spectral Differential Imaging (SDI) data sets, the dispersion solution serves a second purpose by providing the radial $\lambda$/D scaling of the speckles. 

We notice that the strongest gradients in the vlt-sphere dispersion solution coincide with edges of telluric H$_2$O absorption bands (Fig.~\ref{fig:dispersion}, bottom). While the star itself could be considered as a flat continuum source between neighbouring spectral channels, the edges of telluric absorption bands result in strong gradients in the number of photons recorded as a function of wavelength. As a consequence, at the blue-ward edge of a telluric absorption band a channel records more shorter wavelength (``bluer'') than longer wavelength (``redder'') photons. The opposite happens at the red-ward edge of an absorption band. The centroids of satellite spots at the blue edge of an absorption band are thus weighted in favour of short wavelength photons, resulting in a smaller radial separation of opposing spots on the detector. The peak of satellite spots at the red edge of an absorption band are slightly further apart. Thus the vlt-sphere ``dispersion solution'' is not representative of the intrinsic (smooth) response of the IFS AMICI prism to a ``flat spectrum'' source, but representative of the response to a source with the spectral characteristics of the satellite spots.

To correct for the above mentioned effect, one could mask the channels around the water absorption bands and use a cubic relation to fit the position of the satellite spots, which would correct the quadratic dispersion computed by fitting the three (four) diode lasers observed in the wavelength calibration for the $YJ$ ($YH$) IFS modes.

\section{Telluric monitoring and correction \label{sec:data_reduction_sphere_tellurics}}

\begin{figure}
       \begin{center}
        \includegraphics[width=0.46\textwidth]{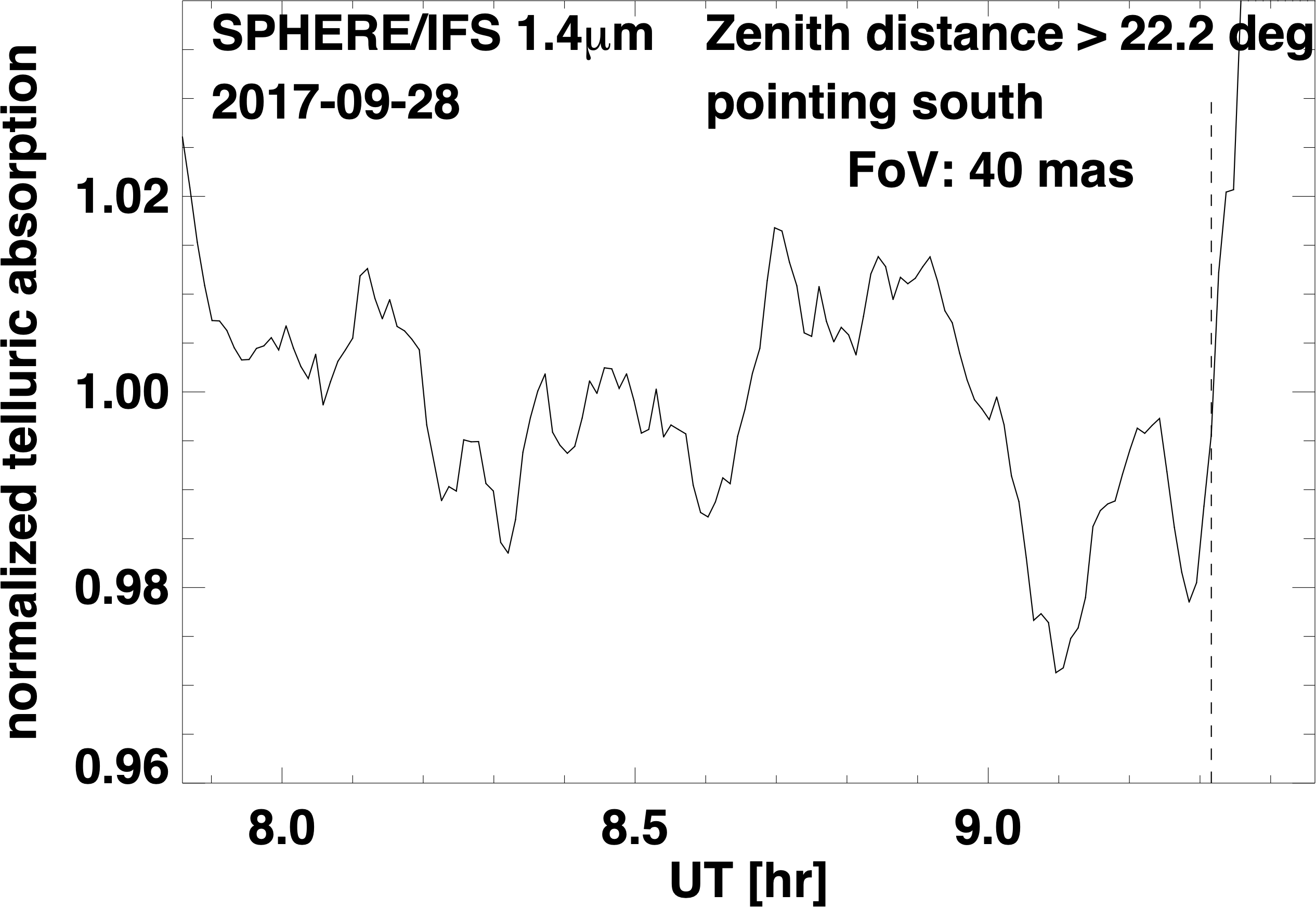}
        \includegraphics[width=0.46\textwidth]{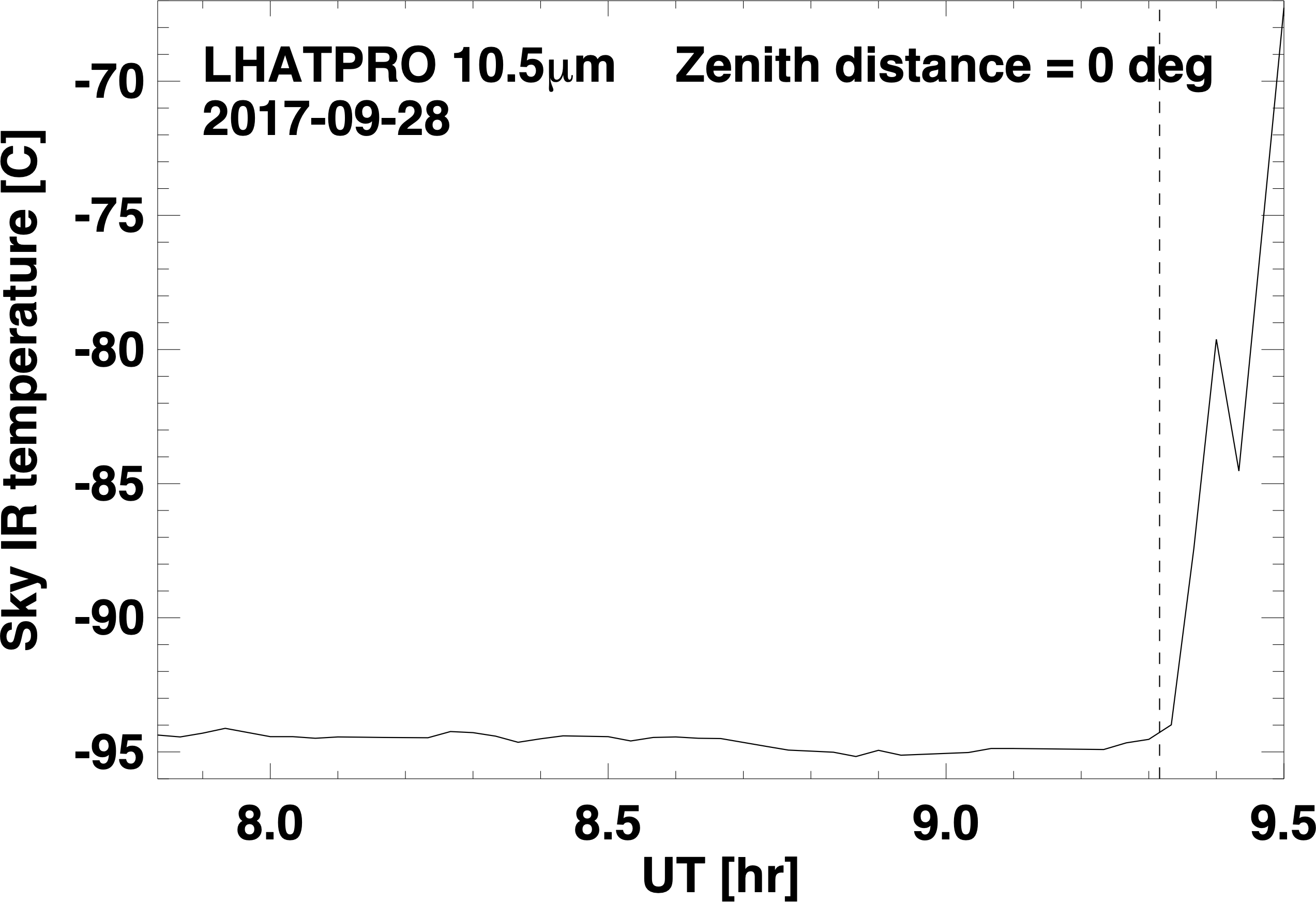}
        \includegraphics[width=0.46\textwidth]{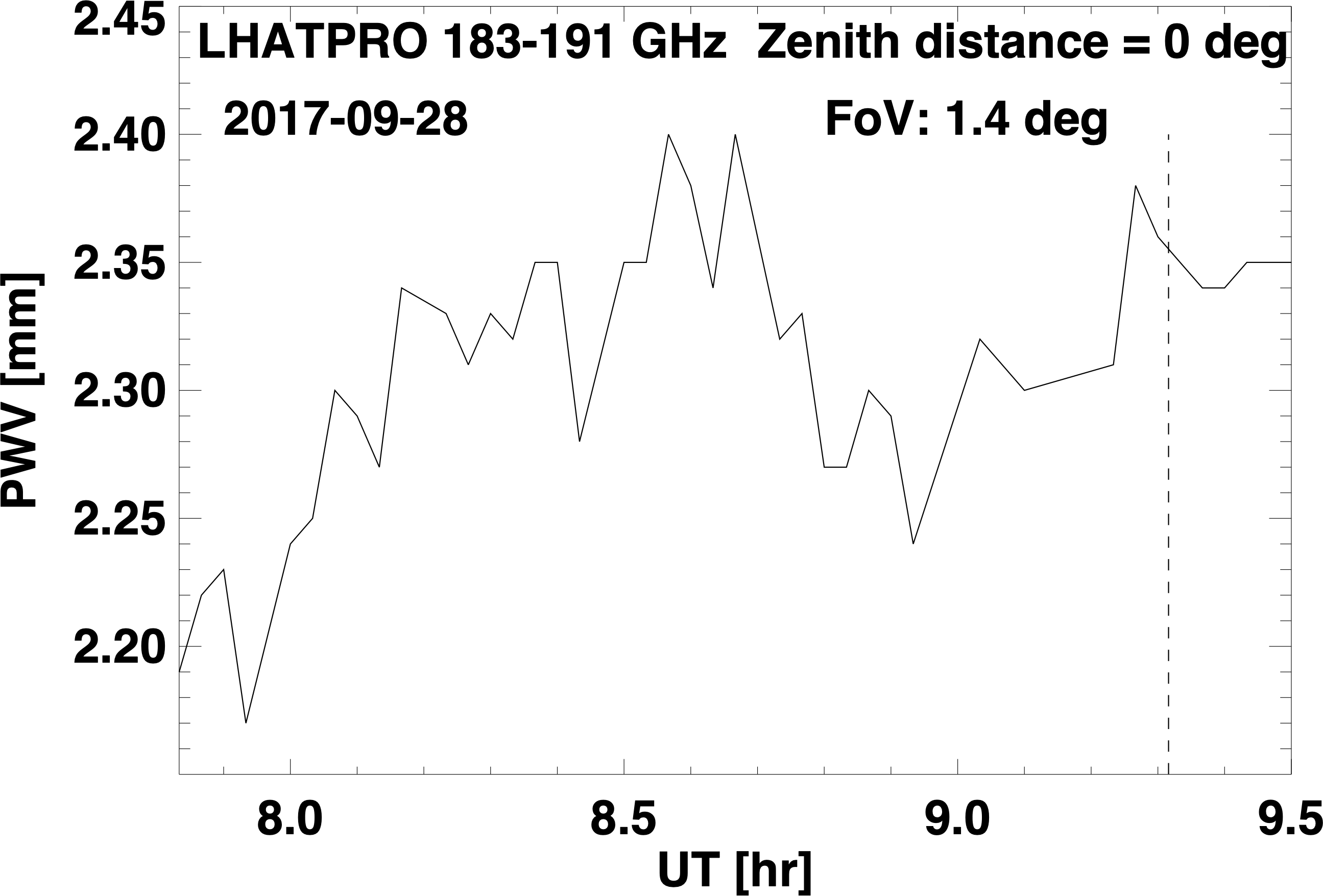}
    \end{center}
      \caption{Top: Telluric absorption as traced by SPHERE/IFS at $\approx 1.4\mu$m. Middle: Sky infrared temperature as traced by LHATPRO. Bottom: PWV as traced by LHATPRO.  The vertical dashed line in all figures marks our cut-off time for the first 140 IFS frames, which form the basis of our spectral analysis, out of a total of 154 IFS data frames.}
    \label{fig:telluric_timeseries}
   \end{figure}

Continuous satellite spots not only provide means for a continuous spatial registration of the star, but also offer a simultaneous monitoring of the (grey) atmospheric transmission, and of variations in the strength of telluric H$_2$O absorption bands. Figure~\ref{fig:telluric_timeseries} visualizes the variations in atmospheric conditions over the duration of the full sequence of 154 frames of the 2017-09-28 IFS data set. In the top panel we plot the inverse of the normalized IFS satellite spot count rates observed in the 1.4\,$\mu$m H$_2$O absorption band. For better comparison with the Paranal atmospheric monitoring data as made available by the ESO archive\footnote{https://archive.eso.org/wdb/wdb/asm/lhatpro$\_$paranal/form}, we smoothed the SPHERE/IFS data to the same coarse time sampling of $\approx$120\,s. The most noticeable feature is the sharp rise in absorption near the end of the sequence due to an incoming cloud layer. In the middle panel of Fig.~\ref{fig:telluric_timeseries} we show the contemporaneous sky infrared temperature as measured by the Low Humidity and Temperature Profiling (LHATPRO) instrument \citep{Querel2014}. The rise in the telluric absorption seen in the IFS data approximately coincides with the increase in the sky infrared temperature due to clouds. In the bottom panel of Fig.~\ref{fig:telluric_timeseries} we show the precipitable water vapour (PWV) measured by LHATPRO. We notice that there is no exact synchronicity between the IFS measurements in the telluric H$_2$O absorption band and the LHATPRO PWV. Some shape similarity of the PWV variations between UT $\approx$ 8.1\,hr and $\approx$ 9.1\,hr, and the IFS variations between UT $\approx$ 8.3\,hr and $\approx$ 9.3\,hr, which might be explained by telluric water vapour fluctuations first crossing the LHATPRO field of view, and $\approx$ 12\,min later the SPHERE/IFS field of view, is most likely coincidental.  
A strict correlation between IFS and LHATPRO telluric measurements is not expected, as they monitor different parts of the sky (SPHERE/IFS was tracking 51\,Eri, and LHATPRO was staring at zenith), and also cover different fields of view (40\,mas for SPHERE/IFS vs. 1.4\,deg for LHATPRO). 

The data stress the importance of a simultaneous monitoring of the telluric absorption along the line of sight for high precision (better than $\pm$2\% for the first 140 frames of the present data set) spectro-photometric observations of exoplanets. This can be accomplished either by employing high spectral resolution, which facilitates the monitoring of individual lines in telluric H$_2$O absorption bands, or -- in the case of low- to medium spectral resolution (R $\lesssim$20 000) observations -- by simultaneous monitoring of the spectro-photometric signal of the host star (employing, e.g., continuous satellite spots). 

\section{Spectral correlation matrix \label{sec:corr_mat}} 

\begin{figure}
   \centering
   \includegraphics[width=0.5\textwidth]{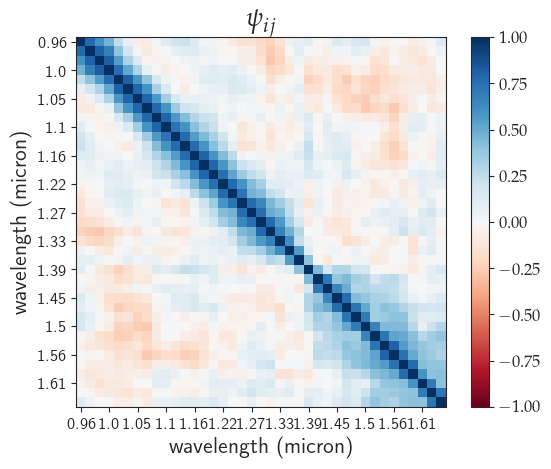}
      \caption{Spectral correlation matrix between each pair of spectral channels in our SPHERE/IFS data.
              }
         \label{fig:correlation}
\end{figure}

The extracted exoplanet spectrum from our SPHERE/IFS data is affected by spectral covariance, which can alter the values of the fitted atmospheric parameters. In order to see by how much our data is affected by this, we followed the methods in \cite{Greco2016} to estimate the average spectral correlation $\psi_{ij}$ within an annulus of 1.5$\lambda/D$ at the separation of the planet, masking out the planet in a 2$\lambda/D$ radius. Where 

\begin{equation}
        \psi_{ij} \equiv \frac{C_{ij}}{\sqrt{C_{ii}C_{jj}}} = \frac{\langle I_i I_j\rangle}{\sqrt{\langle I^2_i \rangle \langle I^2_j \rangle}},
\end{equation}

here, $C$ is the covariance matrix, and $\langle I_i \rangle$ is the average intensity within the annulus at wavelength $\lambda_i$. The covariance matrix is then used to compute the log-likelihood ln $\mathcal{L}$ according to

\begin{equation}
    -2 \ln{\mathcal{L}} \equiv \chi^2 = (S-F)^T C^{-1} (S-F),
\end{equation}

where $S$ is the observed spectrum, and $F$ is the model spectrum. The correlation matrix for our IFS $YH$ spectrum is shown in Fig.~\ref{fig:correlation}.

We ran a ``nominal'' retrieval for our 2018 data with 4000 live points using the covariance matrix to compute the log-likelihood. We observed that the values of the fitted parameters remain within error bars to the ones from the retrieval for which we did not use the covariance matrix. However, the best-fit model has higher and lower flux in the $Y$ and $J$-bands, respectively, compared to the ``nominal'' best-fit model without using the covariance matrix. For this reason, we decided to include the covariance matrix in all our retrievals.

\FloatBarrier

\section{Enforced clouds retrieval and attempt to reproduce previous results\label{sec:repr_samland}}

\begin{figure*}
\sidecaption
    \includegraphics[width=12cm]{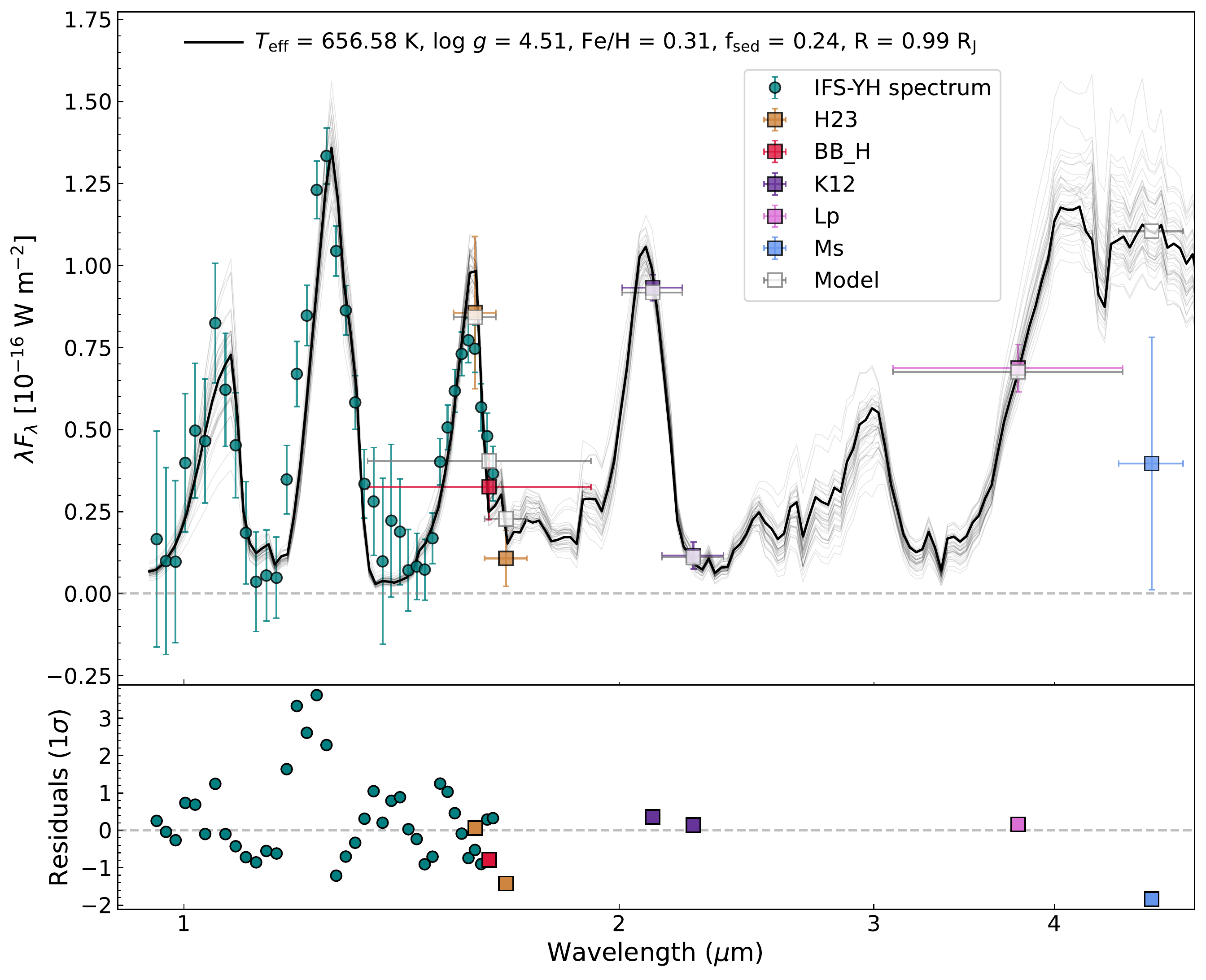}
        \caption{\texttt{petitRADTRANS} best-fit spectrum of 51~Eri~b for the ``enforced clouds'' retrieval run on our new SPHERE spectro-photometric data (teal circles and purple squares) along with the photometric points included in \cite{Samland2017} (shown as squares). The photometric points describe the average flux in the respective filter, the \textit{x}-error bar represents the filter widths. 34 randomly drawn samples from the posterior probability distribution are shown in gray, to show the spread of model parameter combinations to fit the data. Residuals in multiples of 1 $\sigma$ uncertainties of the data are shown below.}
        \label{fig:new_cloudy}
\end{figure*}

\begin{figure*}
   \centering
    \includegraphics[width=0.85\textwidth]{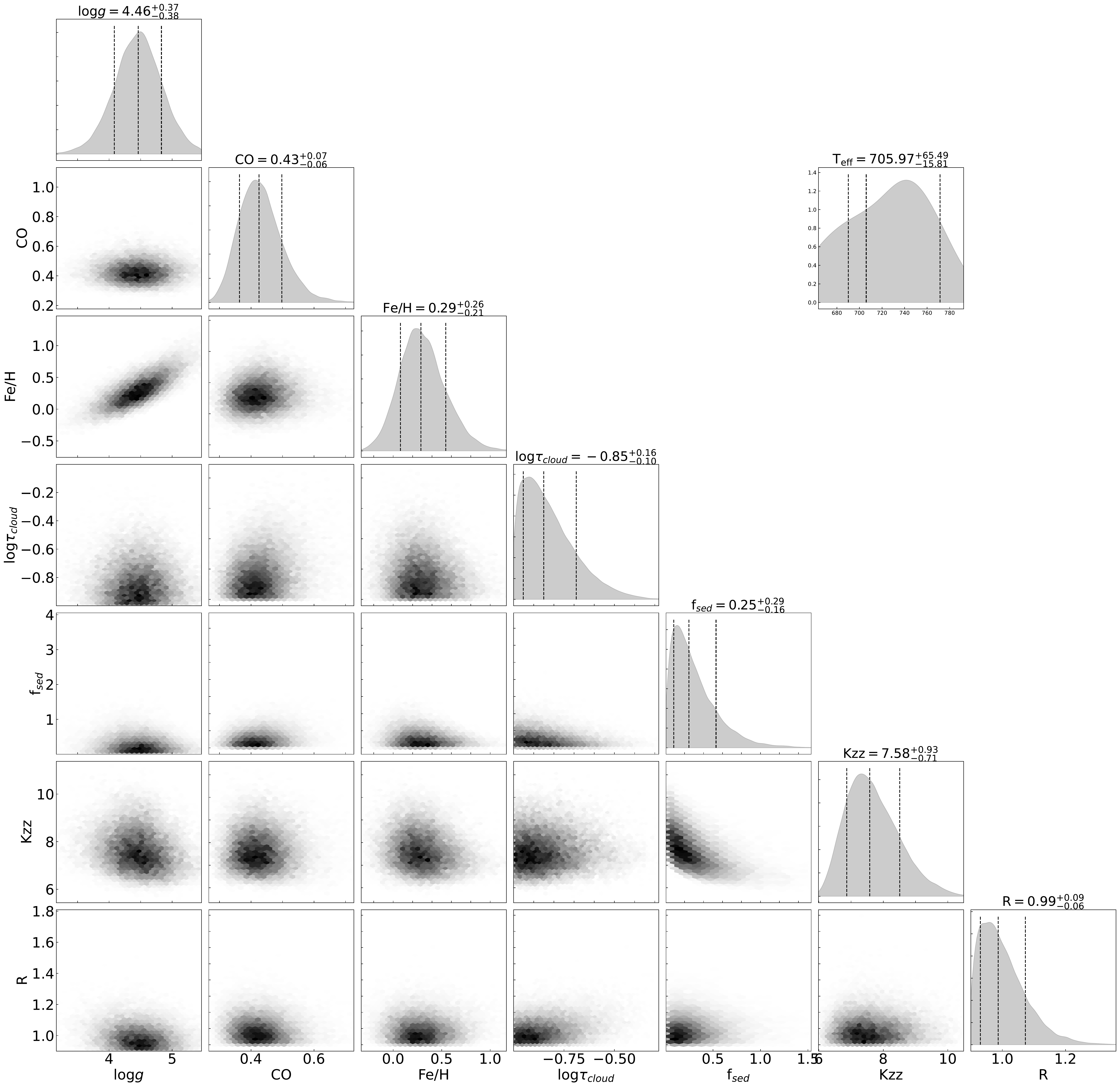}
   \caption{Corner plot of the posterior PDFs of the ``enforced clouds'' retrieval run on the new data set.}
              \label{fig:cloudy_corner}
\end{figure*}

\begin{figure*}
\sidecaption
    \includegraphics[width=12cm]{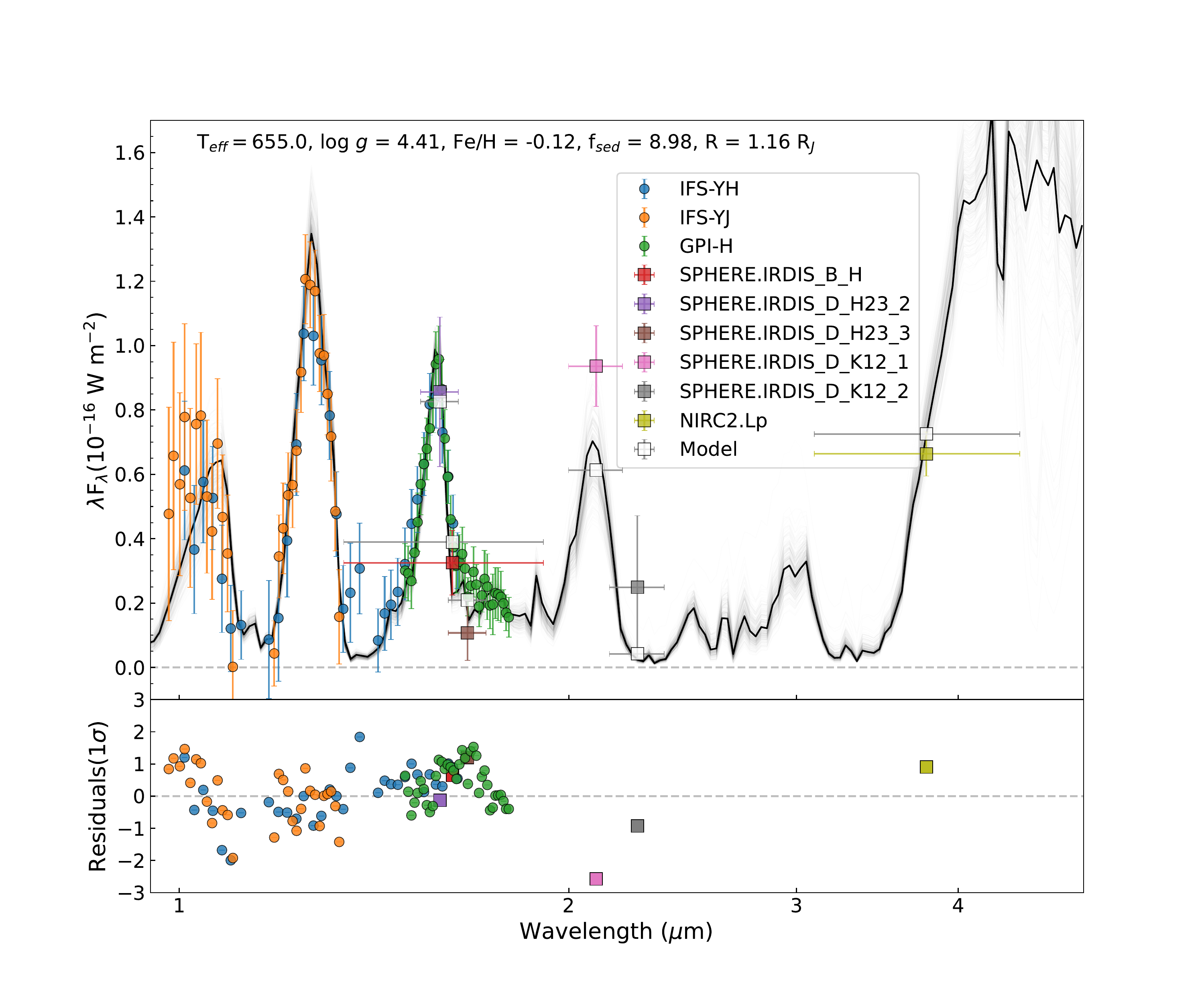}
    \caption{Best fit spectrum (top) of the nominal retrieval run on the original data set from \cite{Samland2017}. This is to be compared to Fig.~11 in \cite{Samland2017}.}
    \label{fig:repr_samland_free_spec}
\end{figure*}

 \begin{figure*}
   \centering
    \includegraphics[width=\textwidth]{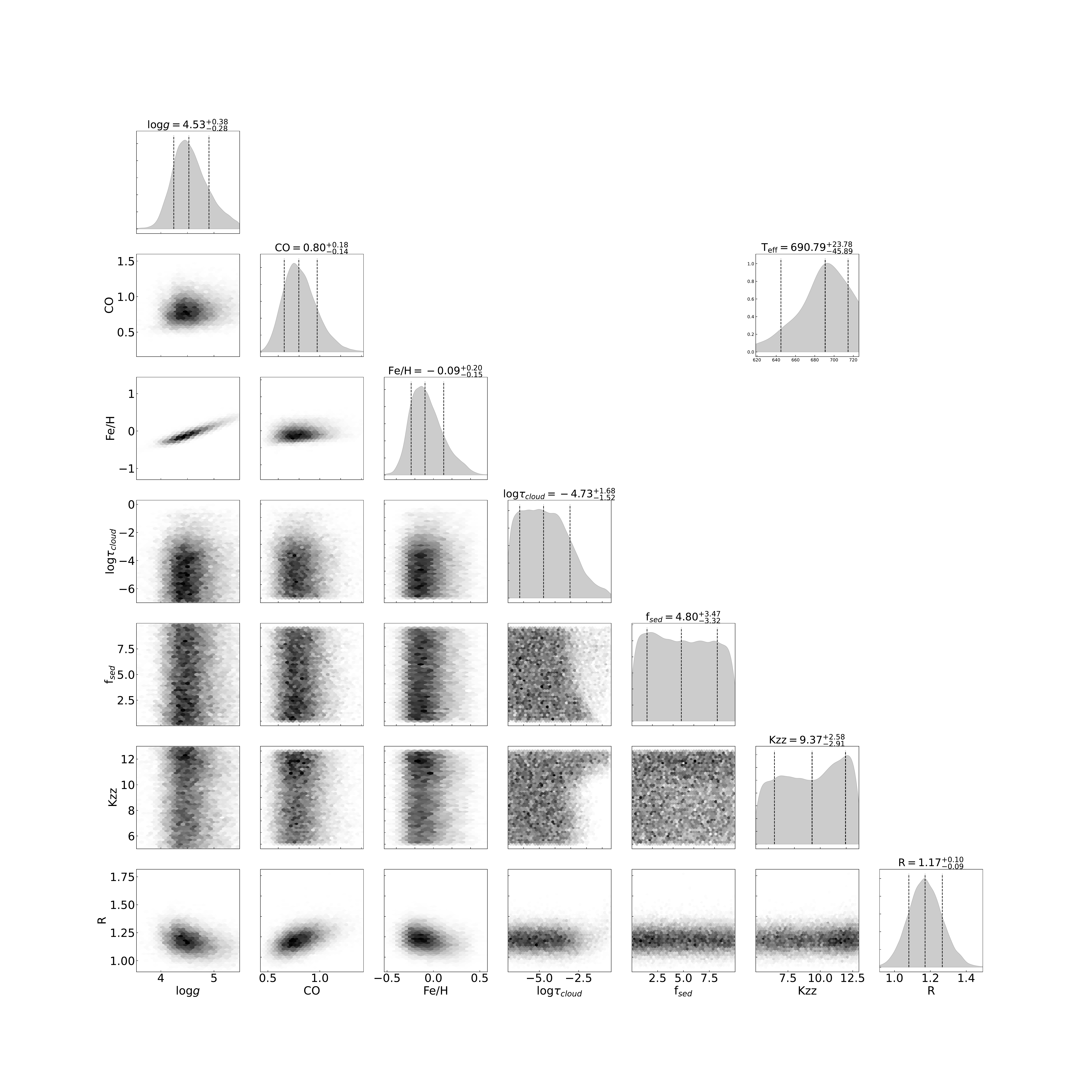}
    \caption{Corner plot of the posterior PDFs of the retrieval run on the original data set from \cite{Samland2017}. This is to be compared to Fig.~12 in \cite{Samland2017}.}
    \label{fig:repr_samland_free_corner}
\end{figure*}

\begin{figure*}
\sidecaption
    \includegraphics[width=12cm]{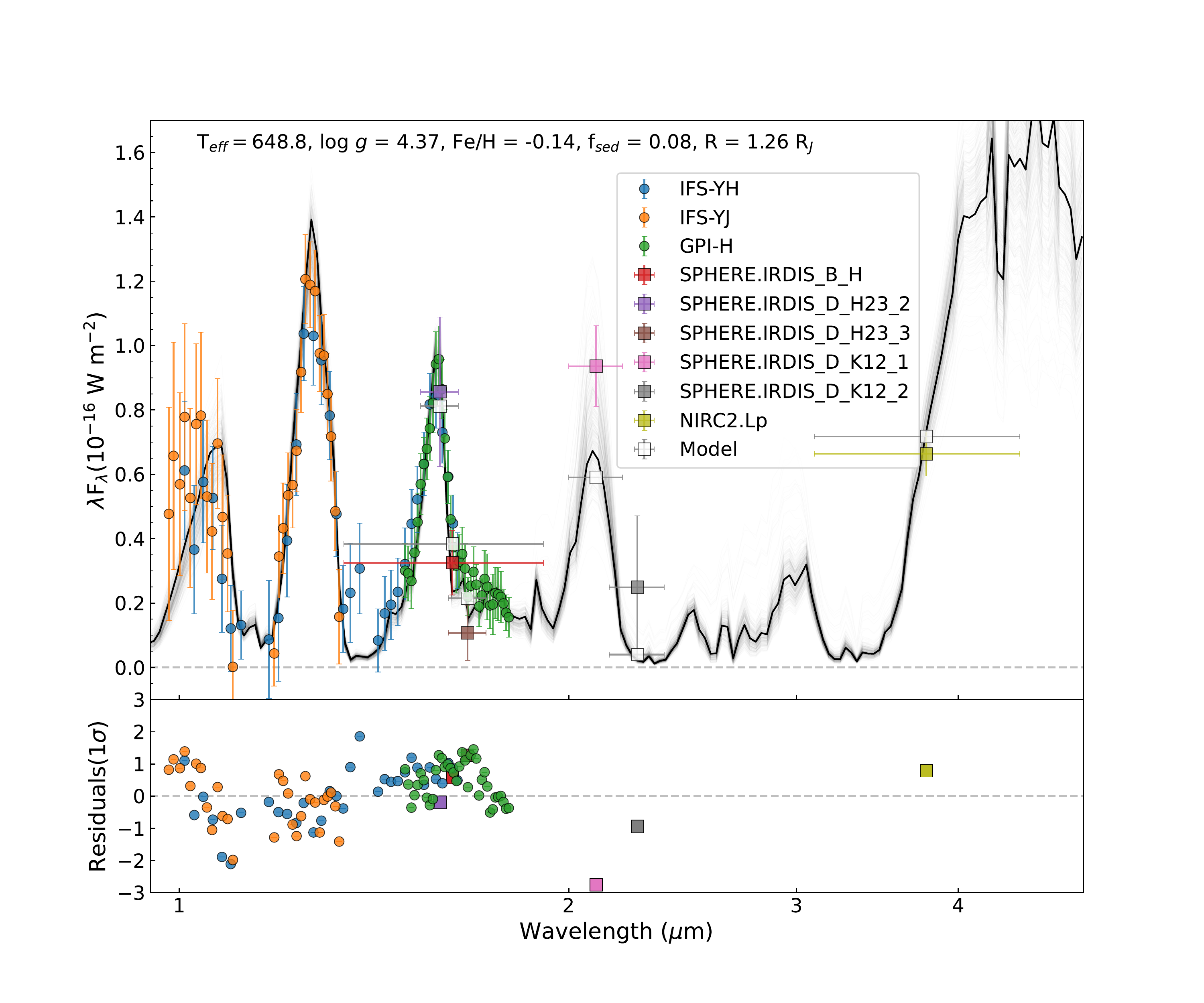}
    \caption{Best fit spectrum of the posterior PDFs  of the retrieval run on the original data set from \cite{Samland2017} when restricting the range of the $\tau_{\mathrm{cloud}}$ prior to positive values. }
    \label{fig:repr_samland_cloudy_spec}
\end{figure*}

\begin{figure*}
   \centering
   \includegraphics[width=\textwidth]{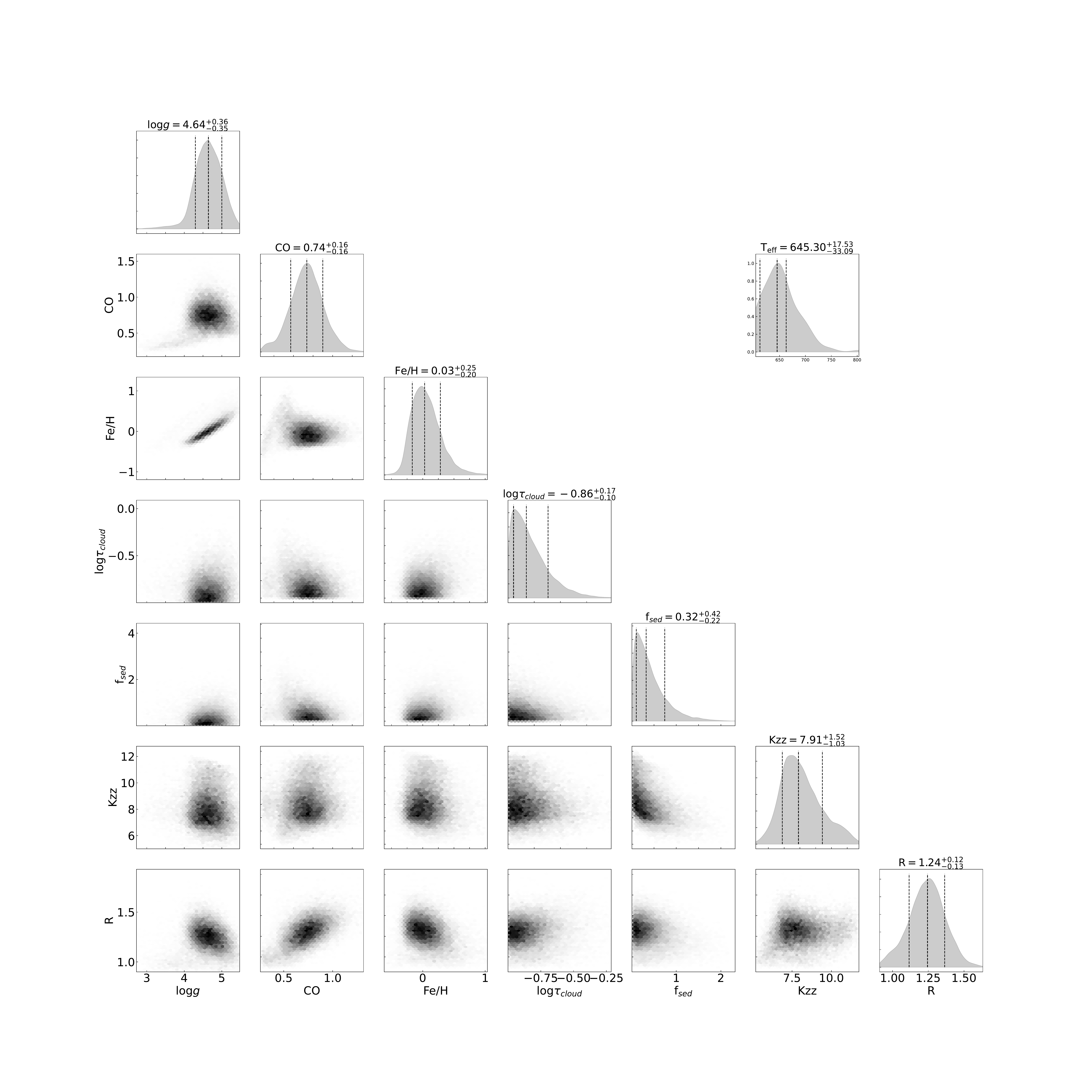}
   
   \caption{Corner plot of the posterior PDFs of the retrieval run on the original data set from \cite{Samland2017} when restricting the range of the $\tau_{\mathrm{cloud}}$ prior to positive values. }
              \label{fig:repr_samland_cloudy_corner}
\end{figure*}

\begin{table*}
\caption[Best fit parameters for the reproduction attempts]{Obtained parameter values from \citetalias{Samland2017} data\label{tab:repr-best}} 
	\centering
	\def\arraystretch{1.5}
	\begin{tabular}{ccccccccccc}
	\hline
    \hline
    	Model & log $g$ & radius & [Fe/H] & CO & f$_{\rm sed}$ & T$_{\rm eff}$ & log$\tau_{\mathrm{cloud}}$ \\
        \hline
        Nominal  &  $4.02^{0.30}_{0.34}$ & $1.00^{+0.08}_{-0.06}$ & $-0.20^{+0.20}_{-0.18}$ & $0.39^{+0.08}_{-0.07}$ & $4.39^{+3.70}_{-3.11}$ & $716^{+24}_{-36}$ & $-5.0^{+1.6}_{-1.3}$ \\
		Enforced clouds  & $4.02^{+0.33}_{-0.34}$ & $1.56^{+0.23}_{-0.27}$ & $-0.06^{+0.20}_{-0.19}$ & $0.42^{+0.05}_{-0.04}$& $0.15^{+0.13}_{-0.08}$ & $510^{+15}_{-14}$ & $0.05^{+0.05}_{-0.03}$ \\  \hline
		Samland17  & $4.26^{+0.24}_{-0.25}$ & $1.11^{+0.16}_{-0.13}$  & $1.03^{+0.10}_{-0.11}$ & -- & $1.26^{+0.36}_{-0.29}$ & $760^{+21}_{-22}$ & -- \\
        \hline
	\end{tabular}
\label{tab:repr_samland}
\end{table*}

Figure~\ref{fig:new_cloudy} shows the best-fit spectrum along with 34 randomly sampled posterior PDFs for the ``enforced clouds'' retrieval using our new data + photometric points. The best-fit parameters are quoted in Table~\ref{tab:results}, and the posterior PDFs are shown in Fig.~\ref{fig:cloudy_corner}, the values quoted for each parameter correspond to the median of the posterior distribution, the uncertainties show the 16th and 84th percentile, representing a 1$\sigma$ uncertainty range. In this way the values are not identical to the ones that produce the overall best fit, which are given at the top of Fig.~\ref{fig:new_cloudy}. Being the median, they also do not necessarily correspond to the most probable value that can be seen from the peak of the marginalized posterior distribution shown in Fig.~\ref{fig:cloudy_corner}.

Figure~\ref{fig:repr_samland_free_spec} shows the best-fit spectrum along with 100 randomly sampled posterior PDFs for the ``nominal'' retrieval using \citetalias{Samland2017}'s data. Figure~\ref{fig:repr_samland_free_corner} shows the corresponding posterior PDFs. The parameters for the best fit spectra are summarized in Table~\ref{tab:repr_samland}. Values and errors quoted in Table~\ref{tab:repr-best} are derived from the equally weighted posterior distribution produced by the Multinest algorithm for each parameter, i.e. marginalized over all parameters except the one in question.
The last line gives the parameters derived by \cite{Samland2017} for their best-fitting ``PTC-C'' model. That latter model implies a cloud fraction of 100\%.
Figures~\ref{fig:repr_samland_cloudy_spec} and \ref{fig:repr_samland_cloudy_corner} are analogous to the above but for the ``enforced clouds'' case.

As can be seen from Table~\ref{tab:repr_samland} we reproduce most of the parameters to within the calculated uncertainties, albeit with the major difference that our atmosphere shows no significant trace of clouds, and our metallicity is sub-stellar whereas \citetalias{Samland2017} found a strongly super-stellar metallicity. 
Be reminded that $\tau_{\mathrm{cloud}}$ denotes the optical depth of the cloud deck at the location where the atmosphere becomes optically thick due to gas opacities alone, i.e. $\tau_{\mathrm{gas}}\approx1$. This implies, that our nominal solution shows essentially no clouds at all ($\tau_{\mathrm{cloud}}\approx10^{-5} @ \tau_{\mathrm{gas}}\approx1$), whereas the enforced clouds solution ($\tau_{\mathrm{cloud}}\approx1 @ \tau_{\mathrm{gas}}\approx1$) requires a rather unphysical gas giant with a radius of 1.56\,R$_{\mathrm{Jup}}$, which for compensation needs to be unusually cool.

\end{appendix}

\end{document}